\documentclass[authoryear,12pt,3p]{elsarticle}

\usepackage{graphicx}
\usepackage{subcaption}

\usepackage{amssymb}

\journal{Journal of Informetrics}

\begin{document}

\begin{frontmatter}

\title{Predicting citation counts based on deep neural network learning techniques}

\author{Ali Abrishami}
\ead{a.abrishami@mail.sbu.ac.ir}

\author{Sadegh Aliakbary\corref{cor1}}
\ead{s\_aliakbary@sbu.ac.ir}
\cortext[cor1]{Corresponding author}

\address{Faculty of Computer Science and Engineering, Shahid Beheshti University G.C, Tehran, Iran}

\begin{abstract}

With the growing number of published scientific papers world-wide, the need to evaluation and quality assessment methods for research papers is increasing. Scientific fields such as scientometrics, informetrics, and bibliometrics establish quantified analysis methods and measurements for evaluating scientific papers. In this area, an important problem is to predict the future influence of a published paper. Particularly, early discrimination between influential papers and insignificant papers may find important applications. In this regard, one of the most important metrics is the number of citations to the paper, since this metric is widely utilized in the evaluation of scientific publications and moreover, it serves as the basis for many other metrics such as h-index. In this paper, we propose a novel method for predicting long-term citations of a paper based on the number of its citations in the first few years after publication. In order to train a citation count prediction model, we employed artificial neural network which is a powerful machine learning tool with recently growing applications in many domains including image and text processing. The empirical experiments show that our proposed method outperforms state-of-the-art methods with respect to the prediction accuracy in both yearly and total prediction of the number of citations.

\end{abstract}


\end{frontmatter}



\section{Introduction}
\label{sec:intro}

Nowadays, many researchers are working on scientific projects world-wide and writing research papers. As a result, many papers are being published everyday with different scientific qualities and impacts. Therefore, the need to evaluating published papers and assessing their quality is overwhelming. A variety of criteria exist in the literature for evaluating the quality of a scientific paper, but one of the most important evaluation metrics is the number of citations to the considered paper. The citation count is a significant indicator since it is widely used for measuring the impact of a paper \citep{garfield1998use, Moed:2005:CAR:1077093, doi:10.1108/eb026940} and moreover, it has been used as the basis for many other metrics such as h-index \citep{Hirsch16569}, impact factor \citep{doi:10.1001/jama.295.1.90}, i-10 index, and other evaluation metrics for journals, conferences, researchers, and research institutes \citep{Wildgaard2014,Moed2012}.

We consider the problem of predicting the citation count of a scientific paper. This problem has many applications in different domains. With the increasing amount of published papers, researchers need to recognize the more influential papers in advance, so that they can plan their research direction \citep{Yan:2012:BSS:2232817.2232831, AMJAD2017307}. Moreover, by predicting the citation count of a paper, we can evaluate the future impact of the paper authors, with potential applications in hiring researchers and faculties, and granting awards and funds. Various efforts exist in the literature for gaining such insights about the future impact of researchers \citep{Chan2018, Havemann2015, Revesz:2015:DMC:2790755.2790763, FIALA20171044}.

In this paper, we propose a method for predicting the citation count of a scientific paper based on its citations during the early years of publication. In other words, the proposed method takes the citation count of a paper in a few years after its publication (namely three to five years), and then predicts its citations in a more long-term period (e.g. from 5th to 15th year after publication). In this problem, we do not investigate other sources of information such as author features, journal properties, and the content (text) of the paper. Although some existing works include more sources of information as the inputs of the citation prediction method, we limit the inputs as simple as the citation pattern of the early publication year, in order to keep the problem definition simple, general, and also applicable in other domains. This problem has already gained attention in the literature, and various methods are proposed to solve this problem \citep{CAO2016471, Wang127}.

It is worth noting that citation prediction is not a trivial task, since scientific papers show different patterns of citations. For example, some papers remain unnoticed for many years and then, attract a lot of attention (this phenomenon is called "sleeping beauty in science") \citep{Li2016, Ke7426, vanRaan2004}. Some other publications gradually gain less citations due to emergence of new methods or losing applicability. Consequently, a single rule or a simple model can not effectively predict the future citations of a paper, and more powerful methods are needed.

In the terminology of machine learning methods, we considered the citation prediction as a regression learning problem, and then we utilized artificial neural networks as a powerful model for learning the prediction task. Based on the citation patterns of many existing scientific papers, the proposed neural network is trained to predict the citation count of papers in the future. Artificial neural networks are inspired by the human brain networks, and has found many successful applications in regression and classification learning. In recent years, neural networks (particularly deep networks) have been effectively applied in various problems such as voice recognition \citep{6296526, 5740583}, object recognition \citep{Schmidhuber:2012:MDN:2354409.2354694, NIPS2012_4824}, image processing \citep{Wan:2014:DLC:2647868.2654948, He_2016_CVPR}, and text processing \citep{Severyn:2015:LRS:2766462.2767738, Sutskever:2011:GTR:3104482.3104610}. We designed a customized recurrent neural network which is appropriate for learning the sequence of the citations. We have also run comprehensive experiments to show the effectiveness of the proposed method over the existing state-of-the-art baseline methods.

The rest of this paper is organized as follows: In section \ref{sec:related_works} we review the related work. In section \ref{sec:proposedmethod} we formulate the problem and present the proposed method. Section \ref{sec:dataset} describes the prepared dataset which is utilized in our experiments. In section \ref{sec:evaluations} we evaluate the proposed method and compare it with the state-of-the-art baselines. Finally, we conclude the paper in section \ref{sec:conclusion} and describe the future works.

\section{Related Work}
\label{sec:related_works}

Many studies exist in the literature for predicting the impact and success of scientific works. The existing works aim different goals such as citation count prediction for scientific papers \citep{Dong:2015:TPI:2684822.2685314, CAO2016471, Wang127, Yu2014, Yan:2011:CCP:2063576.2063757, 10.1371/journal.pone.0049246, DBLP:journals/corr/DongJC16, Yan:2012:BSS:2232817.2232831, 10.1007/978-3-540-75530-2_10, 10.7717/peerj.4564, Pobiedina2016, BORNMANN2014175}, predicting highly cited papers \citep{10.1007/978-3-642-40319-4_2,0295-5075-105-2-28002,Sarigol2014}, predicting h-index of the researchers \citep{Ayaz2018, DBLP:journals/corr/DongJC16, 6c5266f2df3b4be1bbf055b103a081df}, and predicting the impact factor of scientific journals \citep{ketcham2007predicting,Wu2008,rocha2016journal}.

We may categorize the existing works based on their utilized sources of information for scientific impact prediction. In the first category, the graph of the scientific papers is used as the main source of information \citep{Sarigol2014, Pobiedina2016, doi:10.1108/LHT-02-2017-0044, Butun:2017:SLM:3110025.3110160, 10.1007/978-3-642-40319-4_2, Klimek2016}. Sarigol et al. use the co-authorship network of the scientists and the author centrality measures for predicting highly cited papers \citep{Sarigol2014}. Many existing works tackle this problem as a link prediction problem in the citation network of the papers \citep{Pobiedina2016, doi:10.1108/LHT-02-2017-0044, Butun:2017:SLM:3110025.3110160}. McNamara et al. also investigate the paper neighborhood properties in the citation network in order to forecast the highly cited papers \citep{10.1007/978-3-642-40319-4_2}. Klimek et al. construct a bipartite network of papers and words (only the words in the papers abstract), and analyze this network to find the papers with the highest impact potential \citep{Klimek2016}.

The second category of the existing works utilize information that are available right after the publication of the papers. This information includes the content of the paper (paper text), the publication venue (e.g., the journal or conference), information related to the authors, the subject (research area) of the paper, and the references. Such data which are available right after the publication of a paper, may help gain insight about the impact of the paper in the future. For example, \cite{DBLP:journals/corr/DongJC16} propose to use the following six information sources for predicting whether a paper will increase the h-index of the author in a five-year period after publication: author, topic, references, publication venue, the social network, and the temporal attributes.
\cite{10.1371/journal.pone.0049246} utilizes the information related to the paper authors such as published papers count, the research experience years count, the average annual citations count, and h-index, in order to predict the total author citation count in $k$ successive years. Some methods also utilize the topic of the paper (e.g. its rank and freshness), the authors properties (e.g. rank and h-index), and the publication venue for predicting the paper citation count in the future \citep{Yan:2012:BSS:2232817.2232831,Yan:2011:CCP:2063576.2063757}. Yu et al. utilize 24 features including information about the authors and the publication venue \citep{Yu2014}. Castillo et al. utilize information about the past publications of the authors and the co-authorship network to predict the citation count in the first few years after publication \citep{10.1007/978-3-540-75530-2_10}. Bornmann et al. utilize the journal impact factor, number of the authors, and number of the references \citep{BORNMANN2014175}. 


The third category of information which is utilized in existing works includes data gathered after the publication of the paper. For example, \cite{10.7717/peerj.4564} investigate the popularity of a paper in the web (e.g., in the news and Wikipedia) and social networks (such as Twitter and Facebook), and then show a correlation between this popularity and the amount of citation to the paper. As the main class of interest, many existing works utilize short-term citation count of a paper for predicting its long-term citations \citep{0295-5075-105-2-28002, Wang127, CAO2016471, Yu2014}. The citation count of the early years after publication is an important feature which may contribute to improve the accuracy of citation count prediction methods \citep{Kosteas2018,BORNMANN2014175}. For example, \cite{0295-5075-105-2-28002} predicts highly cited papers based on the short-term citation count of the papers and the computation of z-score of the citations of the papers of the same research field. In a significant research, Wang et al. designed a general relationship (formula) which predicts the citation count of a paper in the years after publication \citep{Wang127}. In this method, the prediction formula is fitted to a specific paper by tuning the formula parameters based on the citation count of the paper in the early years of its publication. This work, which served as a leading research for citation count prediction, followed a manual designed formula which is developed by human experts. But in recent years, artificial intelligence and machine learning methods are widely utilized to replace human-inspired constant models. For example, Cao et al. predict citation count of a paper using the citation pattern of the most similar existing papers according to the citations of the early years after publication \citep{CAO2016471}. This method first clusters the papers according to their similarity in short-term citation patterns, and then predicts the citation count based on the average and the centroid of the clusters. This method outperforms state of the art related works and thus is considered as one of the main baselines in our experimental evaluations.

It is worth noting that three described categories overlap, i.e., some methods belong to more than one category. For example, some existing works utilize both short-term citation counts and the author properties \citep{Yu2014, 10.1007/978-3-540-75530-2_10} and therefore, they belong to both the second and the third illustrated categories. 

In this paper, we only consider the short-term citation count (no other features) for predicting the long-term citations of scientific papers. Significant recent studies have defined the same problem statement with the same input information (i.e., only short-term citations pattern) \citep{CAO2016471, Wang127}. This is an important problem since it does not utilize extra information such as the paper text or the authors background and therefore, the results are applicable when the input data is limited.

\section{Methodology}
\label{sec:proposedmethod}
In this section, we formulate the problem and the assumptions and then, we describe our proposed method along with the details of the employed techniques.

\subsection{Problem Statement}
Suppose that the target paper has received $c_{0}$, $c_{1}$, ... , $c_{n}$ citations respectively in the years after its publication. In other words, $c_{i}$ shows the citation count of the paper in the $i$th year after publication. Assume that we know $c_{0}$, $c_{1}$, ... , $c_{k}$ and we want to predict $c_{k+1}$, $c_{k+2}$, ..., $c_{n}$ for a paper ($k<n$). In other words, the problem is to predict the citation count of a paper until the $n$th year of its publication when we already know its citation count only for the first $k+1$ years (\textit{0}th to \textit{k}th year) after its publication. As we described in the Section \ref{sec:related_works}, we only consider the citations of the first $k+1$ years as the input of the algorithm, and no other information (such as author properties or journal attributes) is utilized. The actual citation count for the $i$th year is called $c_{i}$, and the predicted citation count for the $i$th year is called $\hat{c}_{i}$. Moreover, we define $C$ as the total citations of the paper from ${(k+1)}^{th}$ to $n^{th}$ year of publication, and $\hat{C}$ as the corresponding total predicted citations of the same period, which are described in Equations \ref{eq:C} and \ref{eq:C_hat} respectively.

\begin{equation} \label{eq:C}
	C = \sum_{i=k+1}^n{{c}_{i}}
\end{equation}

\begin{equation} \label{eq:C_hat}
	\hat{C} = \sum_{i=k+1}^n{\hat{c}_{i}}
\end{equation}

A citation prediction method is evaluated according to the accuracy of both $\hat{C}$ and $\hat{c_{i}}$ values. This is because an effective prediction method should estimate both yearly citations and total citations of a paper in the considered time period. Consequently, the problem is to minimize the error of $\hat{C}$ and $\hat{c_{i}}$ values, i.e., $\hat{c_{i}}$ values should be as close and correlated as possible to $c_{i}$ values, and so for $\hat{C}$ and $C$ quantities. Table \ref{table:symbols} summarizes the defined symbols.

\begin{table}
	\begin{center}
		\caption{Table of Symbols}
		\label{table:symbols}
		\small	
		\begin{tabular}{|c|p{0.97\textwidth}|}
			
			\hline
			$c_i$ & Citation count of the target paper in its $i$th year of publication 
			\\			
			$\hat{c_i}$ & The predicted citation count of the target paper in its $i$th year of publication
			\\
			$k$ & The number of years after publication in which the citation counts are known
			\\
			$n$ & The number of years after publication in which the citation count should be predicted
			\\
			$C$ & Total citations of the target paper from the ${(k+1)}^{th}$ to $n^{th}$ years after its publication
			\\
			$\hat{C}$ & Total predicted citations of the target paper in the period of ${(k+1)}^{th}$ to $n^{th}$ years after its publication
			\\
			\hline
		\end{tabular}
	\end{center}
\end{table}

\subsection{Citation Count Prediction}

We utilize machine learning methods in order to build a model that learns to predict citation count of a paper in the future based on its citation history. The model should predict the citation count as a non-negative integer number and therefore, the defined problem is regarded as a ``regression problem'' in the context of machine learning methods. Artificial neural network is one of the most powerful and effective methods for regression learning and thus, we designed a special neural network as the main component of our proposed method. A neural network is built up of several layers of neurons, and it learns to find a relationship (function) between the inputs and the outputs of the training data. In recent years, neural network has found significant applications in text processing \citep{mikolov2010recurrent, Lai:2015:RCN:2886521.2886636, Severyn:2015:TSA:2766462.2767830}, image processing \citep{DBLP:journals/corr/SimonyanZ14a, 7534740, NIPS2012_4824}, voice processing \citep{6638947}, and many other fields. Particularly, the so called ``deep learning'' techniques are developed in recent years with surprising power for learning complex functions \citep{Goodfellow-et-al-2016}.

A neural network is first trained in the training phase, and then it is used for prediction in the sampling phase. In our proposed method, a neural network is trained which having the citation count history of a paper in its early years of publication as the input, predicts the citation count in the future years as the output. In other words, the proposed neural network learns to predict $\hat{c}_{k+1}$, $\hat{c}_{k+2}$, ..., $\hat{c}_{n}$ values based on $c_{0}$, $c_{1}$, ..., $c_{k}$ values. The neural network is first trained using a dataset of existing papers with known citations history, and then it can be utilized as the citations estimator in the future based on the pattern of the previous citations of the considered paper. As the datasets of training and test data, we utilize many published papers with known citations history in their first $n$ years after publication. We employ some of those papers as the training set in order to train the neural network, and then we utilize the rest of the papers as the test set in order to evaluate the accuracy of the trained neural network.

In the defined citation count prediction problem, both the inputs and the outputs form sequence of consecutive values. In such problems, Recurrent Neural Networks (RNN) \citep{58337, rumelhart1986learning} are known to be effective for learning the sequence of the values and therefore, we utilized RNNs in our proposed method. RNNs are capable of learning tasks in which the inputs conform an inherent sequence. For example, speech to text \citep{6638947}, sentiment analysis \citep{Severyn:2015:TSA:2766462.2767830}, and machine translation \citep{DBLP:journals/corr/BahdanauCB14, DBLP:journals/corr/ChoMGBSB14} methods are frequently trained by RNNs since their input data are inherently sequential (rather than a set of independent features). As an alternative to RNNs, simple feedforward neural networks are unable to effectively realize the sequence nature of the input and therefore, result in less accuracy. RNNs process the input sequence in their inherent order and as the sequence is being processed, a hidden memory is built based on the so-far-visited input data and therefore, the sequence of the input is effectively considered \citep{Chollet:2017:DLP:3203489}. In the defined citation prediction problem, the input sequence includes $k+1$ values ($c_{0}, ... , c_{k}$), and the output sequence includes $n-k$ values ($\hat{c}_{k+1}, ..., \hat{c}_{n}$). Thus, a many-to-many RNN architecture is designed in our proposed method (RNNs are classified into four categories based on the length of the input and output data: 1- one-to-one, 2- one-to-many, 3- many-to-one, and 4- many-to-many).

\begin{figure}
	\centering
	\includegraphics[scale=0.5]{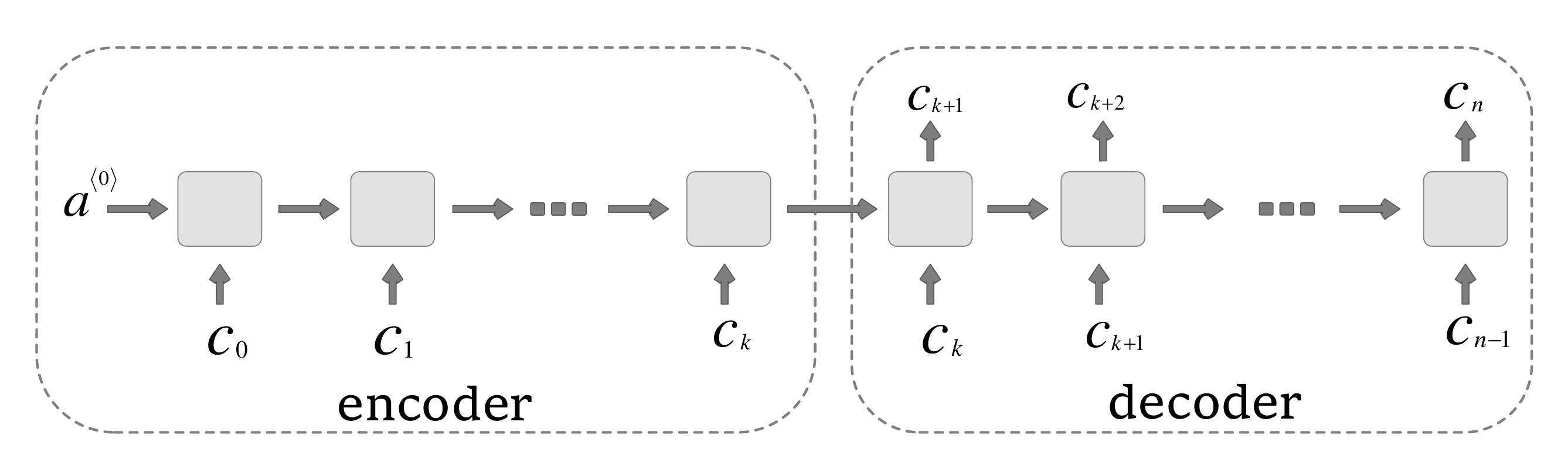}
	\caption{The proposed neural network in the training phase}
	\label{fig:training}
\end{figure}

As the main building block in the architecture of our proposed method, we employed a deep neural network technique called sequence-to-sequence model \citep{NIPS2014_5346} which has already found many successful applications in the literature \citep{NIPS2014_5346, DBLP:journals/corr/ChoMGBSB14}. This technique trains models to convert sequences from one domain to sequences in another domain (e.g., to translate a French sentence to English). The model is composed of two independent neural networks called the ``encoder'' and the ``decoder'' networks. As illustrated in Figure \ref{fig:training}, the inputs are fed to the encoder and the outputs are obtained from the decoder. The encoder network aims to process the input sequence, and create its corresponding representation in another dimensional space. This representation is then forwarded to the decoder network which converts it to the output sequence. In order to predict each output neuron, the decoder utilizes all its input neurons along with the formerly predicted output neurons (see Figure \ref{fig:training}). In this manner, the output neurons are also fed into the network to enable learning the sequence pattern of the input data. 

The proposed neural network should be trained in order to predict the citation count of the papers in the future. In the training phase, the training dataset is utilized in order to optimize the parameters of the neural network. Each tuple of the training set corresponds to the citation information of one published paper, which includes the citation count of the paper from the $0^{th}$ to $k^{th}$ year of its publication as the input, and the actual citation count from $(k+1)^{th}$ to $n^{th}$ year as the output. Therefore, the inputs (citation count values for $k+1$ years) are fed to the encoder and the actual outputs (citation counts from the year $k+1$ to $n$) are presented to the network for training.  As illustrated in Figure \ref{fig:training}, $c_{0}$ to $c_{k}$ are the inputs of the neural network which are fed to the encoder, and $c_{k+1}$ to $c_{n}$ are its outputs which are obtained from the decoder. It is worth noting that the decoder network takes $c_{k}$ as an input in the decoder network in order to predict $c_{k+1}$, and then  $c_{k+1}$ is needed for predicting $c_{k+2}$ and so forth. 

\begin{figure}
	\centering
	\includegraphics[scale=0.5]{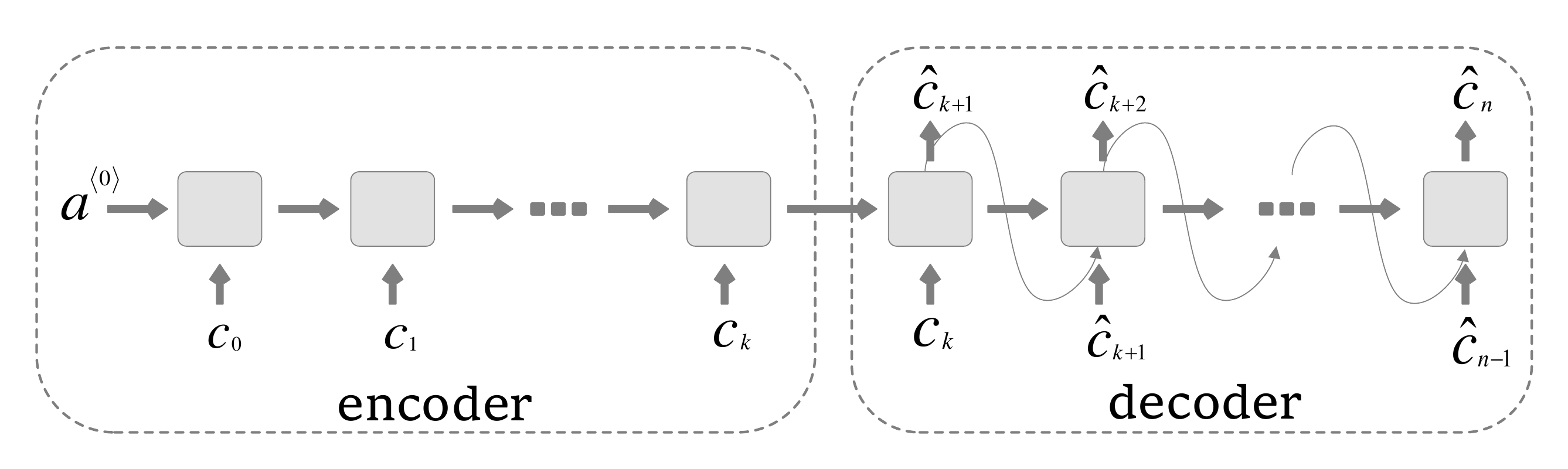}
	\caption{The proposed neural network in the sampling phase}
	\label{fig:sampling}
\end{figure}

After the training phase, the sampling phase aims to predict the citation count for the papers which are not employed in the training phase, having the citation count of their first $k$ years of the paper after publication. Figure \ref{fig:sampling} illustrates the sampling phase, in which the decoder estimates the $c_x$ for $x>k$ values. Since the neural network does not access the actual $c_{x}$ data in the sampling phase, $\hat{c}_x$ predicted values are also fed to the next neuron of the decoder neural network in the sampling phase (this technique is called ``teacher forcing'' \citep{Goodfellow-et-al-2016}). In Figure \ref{fig:training} and Figure \ref{fig:sampling}, $a^{<0>}$ illustrates the initial hidden memory (hidden state) of the encoder network. We have set this initial memory to an all-zero vector which is fed to the encoder network (this initialization is common in deep learning tasks). 

The proposed method, learns to predict the future citation count of a paper based on the history of its early years citations. The employed recurrent neural network with the sequence-to-sequence model technique, enables the method to effectively learn the sequence pattern of the citations, and to predict the future citations more accurately than the baseline methods, as illustrated in section \ref{sec:evaluations}.

\subsection{Implementation Details}
In this section, we overview the implementation details of our experiments in order to make the research results reproducible. Particularly, we describe the implemented neural network and its parameters which resulted in the best outcomes based on our experiments. Such implementation details are worth noting because designing an effective deep neural network is a challenging task for this application which requires investigating many hyper-parameters and design choices. 

We utilized ``Keras'' framework (https://keras.io) which is a well-known and widely used implementation for artificial neural networks and deep learning techniques. In addition, we used the ``SimpleRNN'' module of Keras for implementing the proposed neural network. In order to make the neural network capable of learning a complex function, the recurrent layers include 512 values, which means the encoder network generates a vector output with 512 dimensions. In the decoder sector, a ``Dense'' layer is utilized to generate the output (predicted citation count) namely $\hat{c_{i}}$. We did not normalize the inputs and outputs of the neural network. The ``Rectified Linear Unit'' (ReLU) is used as the activation function in all layers of the neural network, which is defined as $f(x) = max(0,x)$. The ReLU activation function is widely being used in recent years because it results in better precision in many applications \citep{relu}. In order to avoid overfitting, the ``Dropout'' technique \citep{JMLR:v15:srivastava14a} is used with the rate of 0.2 in the RNN layers. This technique enables more epochs (100 epochs in our experiments) in the training phase while minimizing the risk of overfitting. We implemented the RMSProp optimization algorithm, which is an effective method for training neural networks, with the learning rate of $10^{-5}$. Data are fed to the network in batches 256 papers (batch-size = 256). More than half a million of parameters are trained and tuned in the training phase of the proposed method. Table \ref{table:implDetails} summarizes the implementation details of the proposed method.

\begin{table}
	\footnotesize
	\centering
	\caption{The implementation details of the proposed method.}
	\label{table:implDetails}
	\begin{tabular}{|c|c|}
		\hline
		Neural network API & Keras \\ \hline
		RNN module & SimpleRNN \\ \hline
		Output dimensions of the encoder & 512 \\ \hline
		The output layer & Dense layer \\ \hline
		Activation function & ReLU  \\ \hline
		Overfitting prevention technique & Dropout with 0.2 rate  \\ \hline
		Epochs & 100  \\ \hline
		Optimization algorithm & RMSProp  \\ \hline
		Learning rate & $10^{-5}$  \\ \hline
		Batch size & 256  \\ \hline
	\end{tabular}
\end{table}

\subsection{Baseline Methods}

We selected three baselines in order to compare the evaluation results against. First, the ``Mean of Early Years'' method (MEY) is a na\"{\i}ve and simple prediction function which always returns the average of the known citation count in the first years of paper publication as the predicted citation count in the future. Equation \ref{eq:mey} shows the MEY prediction method. For many papers, the citation count in the early years of publication is similar to long-term citations. Although MEY is a simple prediction function, it shows relatively good prediction accuracy in some situations, and outperforming MEY is not a trivial task for citation prediction methods. 

\begin{equation} \label{eq:mey}
	\hat{c}_{k+1} = \hat{c}_{k+2} = ... = \hat{c}_{n} = \frac{\sum_{i=0}^k{c_{i}}}{k+1}
\end{equation}

The second and the third baselines are proposed by \cite{CAO2016471}. As illustrated in Section \ref{sec:related_works}, this research has already outperformed important existing methods, such as \citep{Wang127}, and thus we considered it as one of the main baselines in our experiments. The second baseline, which is called AVR, finds the most similar papers published in the same journal according to the citation count of the paper in the early years of publication and then, utilizes the average citations of those found papers in the subsequent years as the predicted citation count of the paper. The third baseline, which is called GMM, also finds the most similar papers published in the same journal according to the citation count of the early publication years but then, clusters the found papers in three groups using Gaussian Mixture Model (GMM) algorithm and then, the most similar centroid to the target paper is selected. The citation pattern of that centroid is utilized as the prediction function for the citation count of the target paper. Finally, our proposed method is called NNCP (Neural-Network-based Citations Prediction) in the evaluation reports. 

It is also worth noting that AVR and GMM methods utilize an $L$ parameter which indicates the number of nearest neighbor papers in the dataset to the target paper. In other words, AVR and GMM methods first find $L$ papers, $y^{(1)}$, $y^{(2)}$, . . ., $y^{(L)}$, from the database of past papers $D$ with smallest matching error $e_x(y)$, where $x$ is the target paper which we want to predict its citations. In order to optimize $L$ parameter for the baseline methods, we considered the papers published in Nature, and using cross-validation technique, we selected $L=20$ which resulted in the best average accuracy in AVR and GMM methods for $L<100$ values. 

\section{Data}
\label{sec:dataset}
We extracted a dataset of published papers along with their citations from the Web of science (WoS) citation database which is a well-known online scientific citation indexing service. In this dataset, we considered the publications of five prestigious journals: Nature, Science, NEJM (The New England Journal of Medicine), Cell and PNAS (Proceedings of the National Academy of Sciences). The dataset covers the papers which are published from 1980 and before 2003, and it includes 14 years of the citations for each paper (all citations before 2017). The mentioned journals are high-impact publications with a long history. Moreover, one of our main baseline methods \citep{CAO2016471} also included similar journals and therefore, we considered such journals for fair comparison of evaluations. 

In order to separate training and test data, we considered the published papers between 1980 and 1997 as the training set, and the papers published in the five subsequent years (from 1998 to 2002) as the test set for evaluations. Table \ref{table:journals_statistics} illustrates the considered journals in this dataset along with the number of extracted papers from each journal and the size of the training-set and the test-set.

\begin{table}
	\footnotesize
	\centering
	\caption{The selected journals, the number of extracted papers, and the size of training and test sets in our experiments. Data are gathered from Web of Science repository.}
	\label{table:journals_statistics}
	\begin{tabular}{|c|c|c|c|}
		\hline
		Journal name & Paper count in the dataset & Training-set size & Test-set size \\ \hline
		Nature  & 72,797 & 58,068 & 14,729  \\ \hline
		Science & 52,646 & 39,616 & 13,030  \\ \hline
		NEJM    & 39,022 & 31,081 & 7,941  \\ \hline
		Cell    & 10,114 & 8,251 & 1,863  \\ \hline
		PNAS    & 853 & 468 & 385  \\ \hline \hline
		\textbf{\textit{Total}}   & \textit{175,432} & \textit{137,484} & \textit{37,948}  \\ \hline

	\end{tabular}
\end{table}

Our prepared dataset includes the following information for each paper: Paper identifier (a number between one and 175,432), journal identifier, publication year, and the citation counts from the $0^{th}$ to $14^{th}$ year after publication ($c_0$ to $c_{14}$ citation count values). Actually we set $n=14$ (according to Table \ref{table:symbols}, $n$ is the last year after publication in which the citation count is predicted). Table \ref{table:papers_samples} illustrates a small window of the dataset for five sample papers. As it can be seen from the samples, extracting a simple pattern of citations is not trivial, and an intelligent method is necessary for citation pattern prediction.

	\begin{table}
		\caption{Features (citations history) of five sampled papers.}
		\label{table:papers_samples}
		\resizebox{\columnwidth}{!}{%
			
			\begin{tabular}{|c c c c c c c c c c c c c c c c c c|} 
				\hline
				Paper ID & Journal ID & Publication Year & $c_{0}$ & $c_{1}$ & $c_{2}$ & $c_{3}$ & $c_{4}$ & $c_{5}$ & $c_{6}$ & $c_{7}$ & $c_{8}$ & $c_{9}$ & $c_{10}$ & $c_{11}$ & $c_{12}$ & $c_{13}$ & $c_{14}$ \\ [0.5ex] 
				\hline
				P1 & NATURE & 1999 & 1 & 19 & 21 & 22 & 19 & 20 & 17 & 11 & 15 & 13 & 12 & 13 & 14 & 9 & 10 \\  
				\hline
				P2 & NATURE & 1989 & 0 & 1 & 2 & 0 & 0 & 0 & 0 & 0 & 0 & 0 & 0 & 0 & 0 & 0 & 0 \\ 
				\hline
				P3 & NEJM & 1980 & 0 & 3 & 2 & 3 & 2 & 1 & 1 & 1 & 0 & 0 & 0 & 0 & 0 & 0 & 0 \\ 
				\hline
				P4 & CELL & 1994 & 2 & 20 & 29 & 25 & 16 & 23 & 14 & 9 & 9 & 7 & 8 & 4 & 5 & 6 & 9 \\ 
				\hline
				P5 & SCIENCE & 1983 & 0 & 5 & 8 & 11 & 16 & 6 & 2 & 6 & 6 & 1 & 4 & 1 & 1 & 0 & 2 \\ 
				\hline
			\end{tabular}%
		}
	\end{table}

\section{Results}
\label{sec:evaluations}
In this section, we illustrate the results of our comprehensive experiments, empirical evaluations, and a comparison to the baselines.

\subsection{Measurement Criteria}

We utilized two popular criteria in order to evaluate the proposed method and compare it with the baselines. First, Root Mean Square Error (RMSE) and second, the coefficient of determination ($R^2$). RMSE measures the variation of the predicted values to the actual values and thus, lower values of RMSE are desirable. $R^2$-score measures the correlation between the actual and the predicted values. The $R^2$ value is always less than or equal to 1, where $R^2 = 0$ means no correlation and $R^2 = 1$ shows a perfect correlation between the predicted and actual values and therefore, higher values of $R^2$ are desirable. Equations \ref{eq:rmse} and \ref{eq:r2} illustrate RMSE and $R^2$ measurements respectively, in which $y$ is the set of actual values, $\hat{y}$ is the set of the predicted values, and $\overline{y}$ is the average of $y_i$ values.
	
	\begin{equation} \label{eq:rmse}
	RMSE(y, \hat{y}) =
	\sqrt{ \frac{1}{N_{samples}} \sum_{i=1}^{N_{samples}} (y_{i} - \hat{y_{i}})^2}
	\end{equation}
	
	\begin{equation} \label{eq:r2}
	R^{2} (y , \hat{y}) = 1 - \frac {\sum_{i=1}^{N_{samples}} (y_{i} - \hat{y_{i}})^2} {\sum_{i=1}^{N_{samples}} (y_{i} - \overline{y})^2}
	\end{equation}
	
\subsection{Evaluation Results}
We set up different experiments in order to evaluate the proposed method. According to the training-set and the test-set (described in Section \ref{sec:dataset}), we computed the accuracy of the proposed method and the baselines according to two criteria of RMSE and $R^2$. 

\begin{figure}
	
	\begin{subfigure}{0.32\textwidth}
		\includegraphics[width=\linewidth]{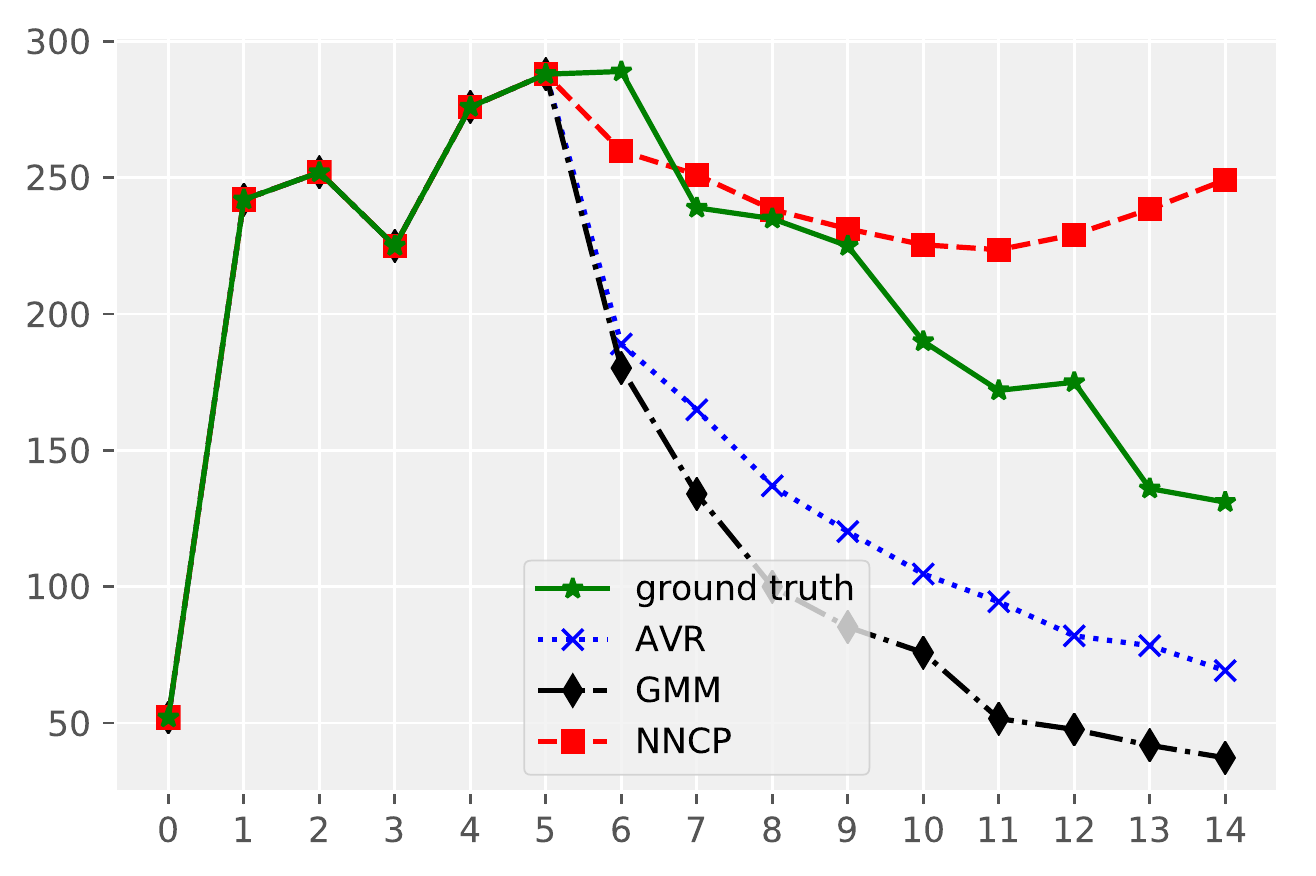}
		\caption{sample paper from Nature}
	\end{subfigure}
	\begin{subfigure}{0.32\textwidth}
		\includegraphics[width=\linewidth]{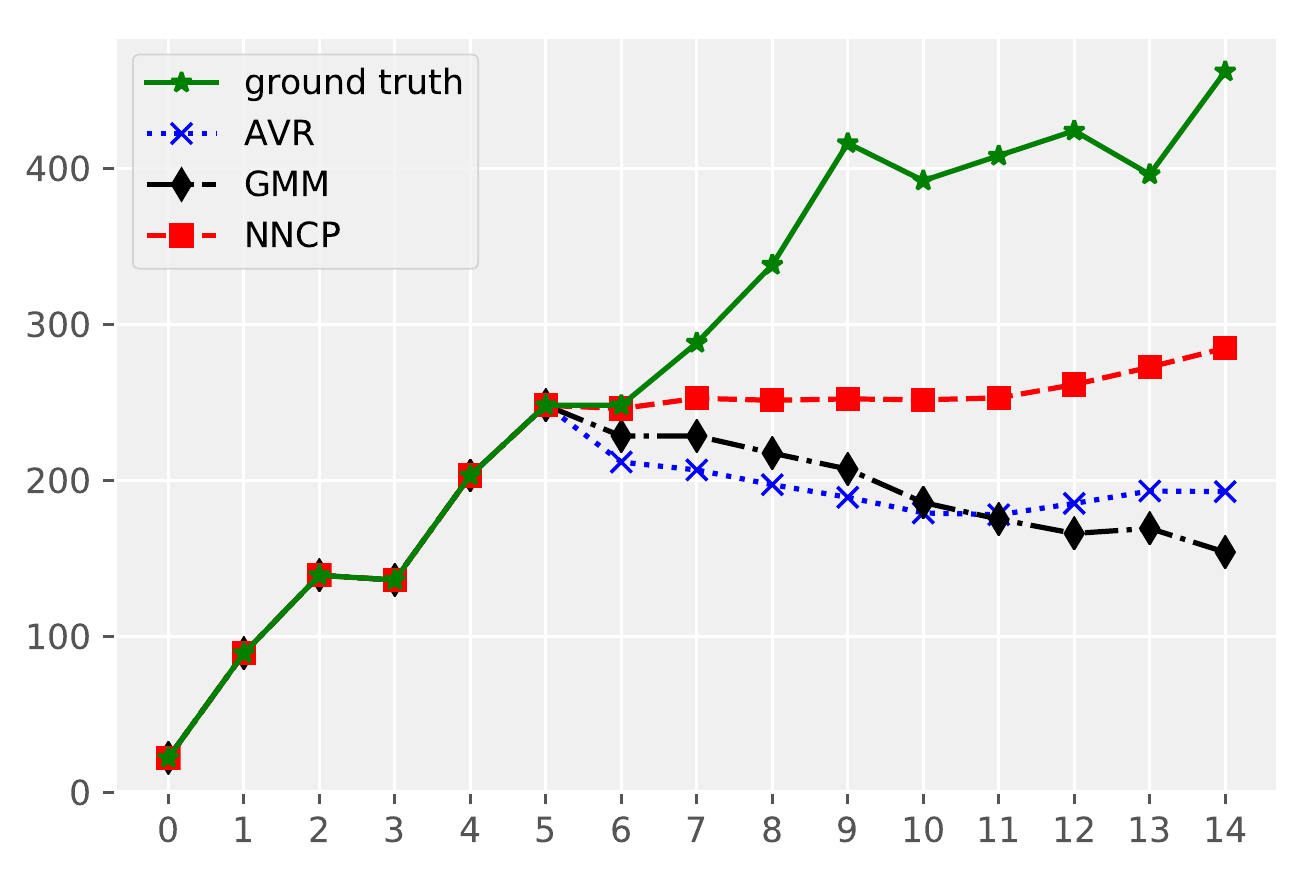}
		\caption{sample paper from Nature}
	\end{subfigure}
	\begin{subfigure}{0.32\textwidth}
		\includegraphics[width=\linewidth]{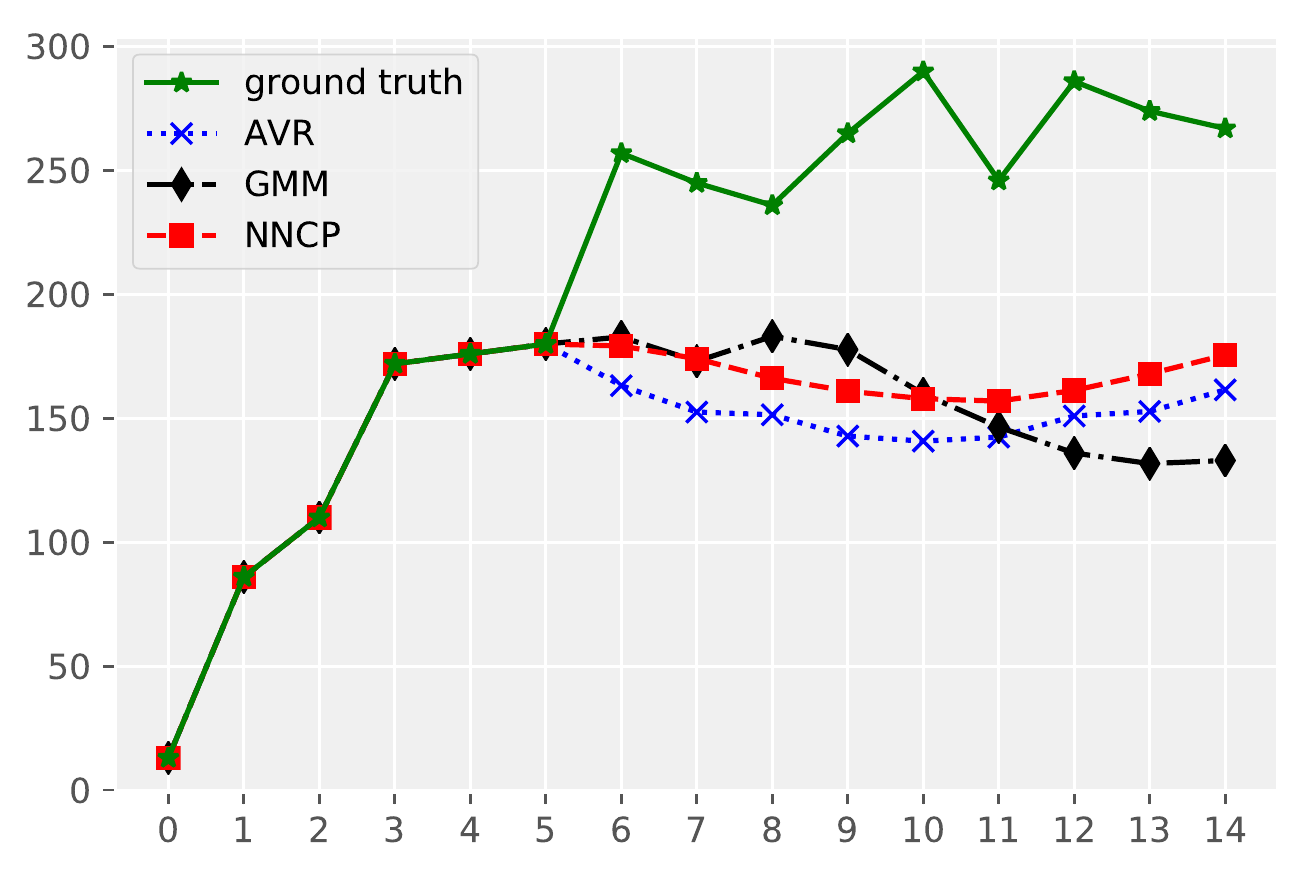}
		\caption{sample paper from Nature}
	\end{subfigure}
	
	\begin{subfigure}{0.32\textwidth}
		\includegraphics[width=\linewidth]{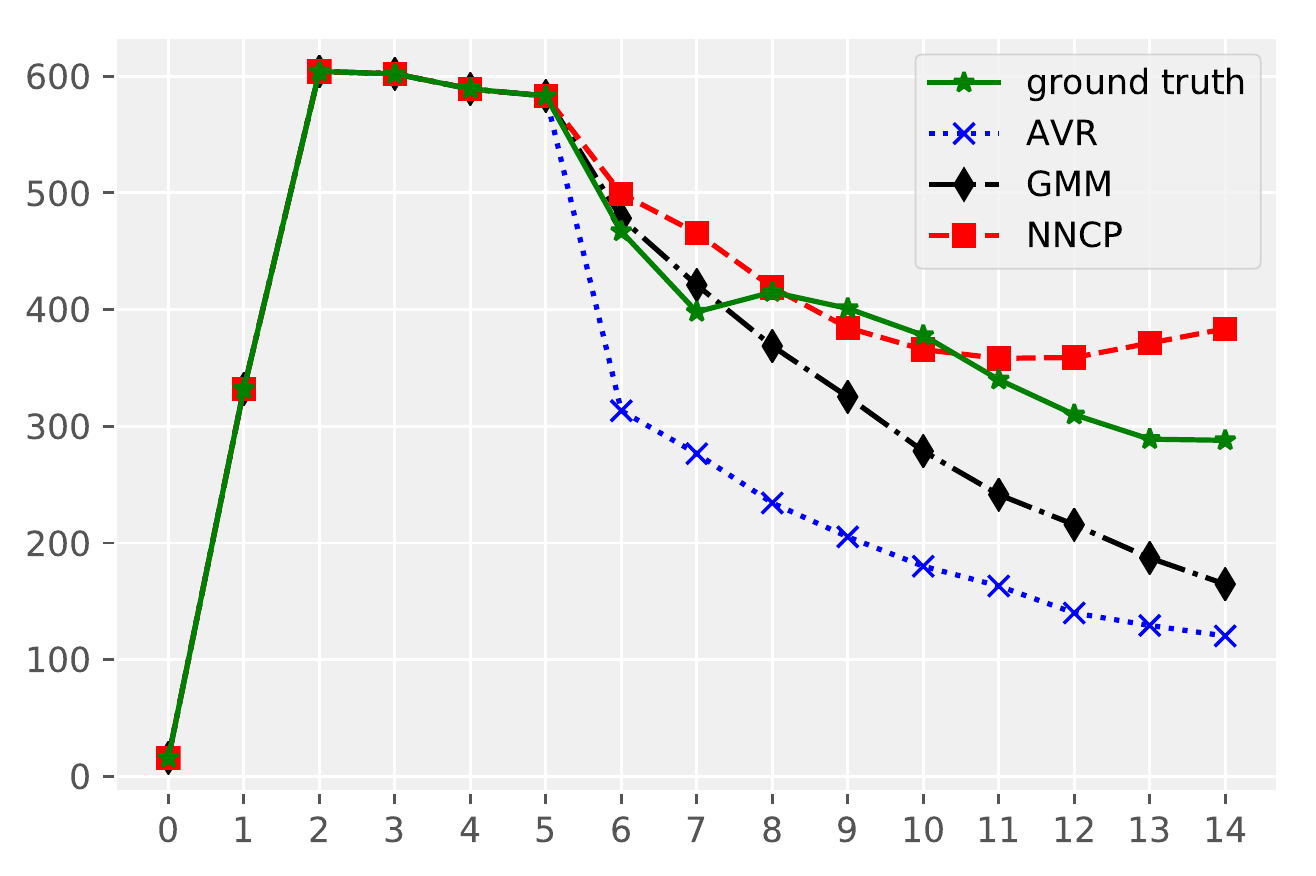}
		\caption{sample paper from Science}
	\end{subfigure}
	\begin{subfigure}{0.32\textwidth}
		\includegraphics[width=\linewidth]{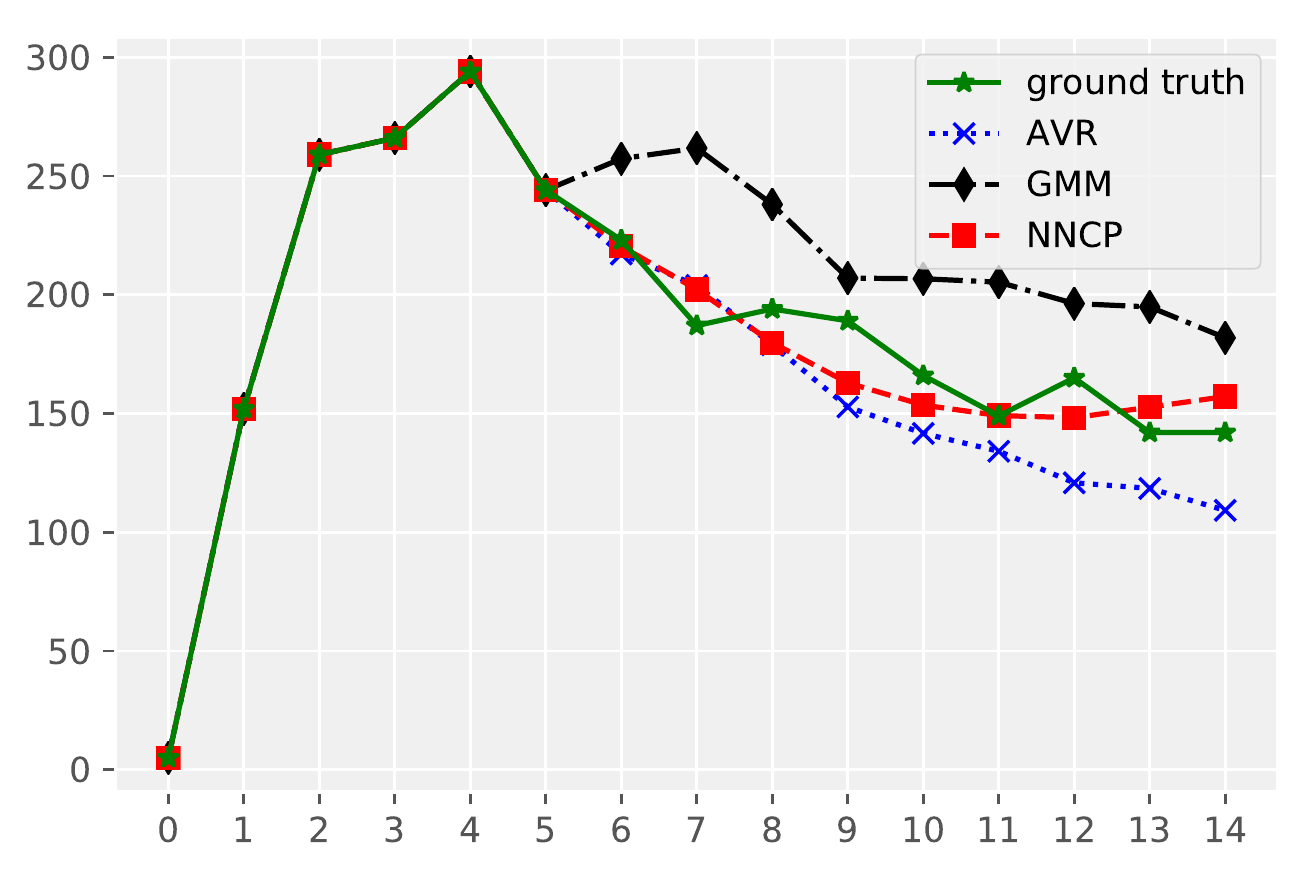}
		\caption{sample paper from Science}
	\end{subfigure}
	\begin{subfigure}{0.32\textwidth}
		\includegraphics[width=\linewidth]{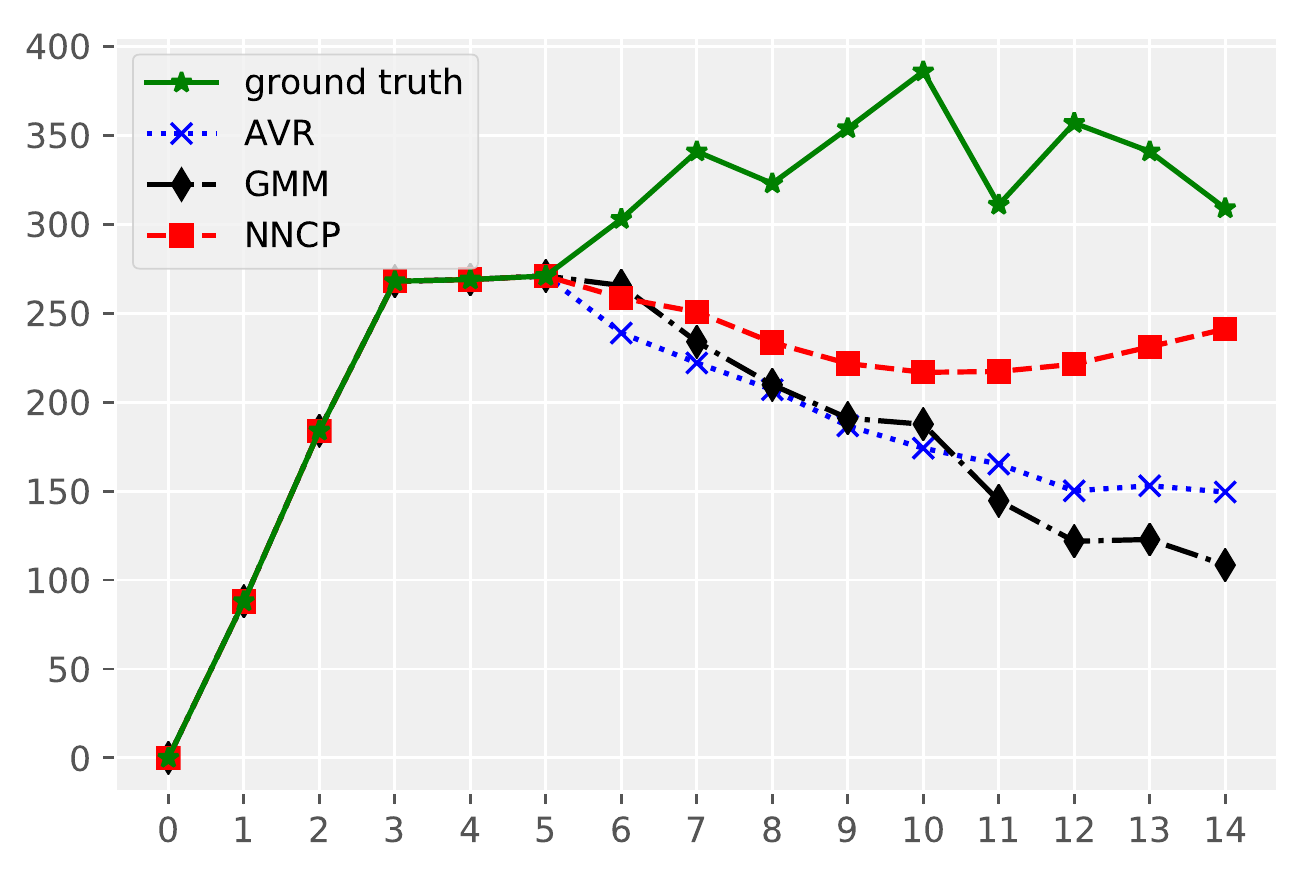}
		\caption{sample paper from Science}
	\end{subfigure}
	
	\begin{subfigure}{0.32\textwidth}
		\includegraphics[width=\linewidth]{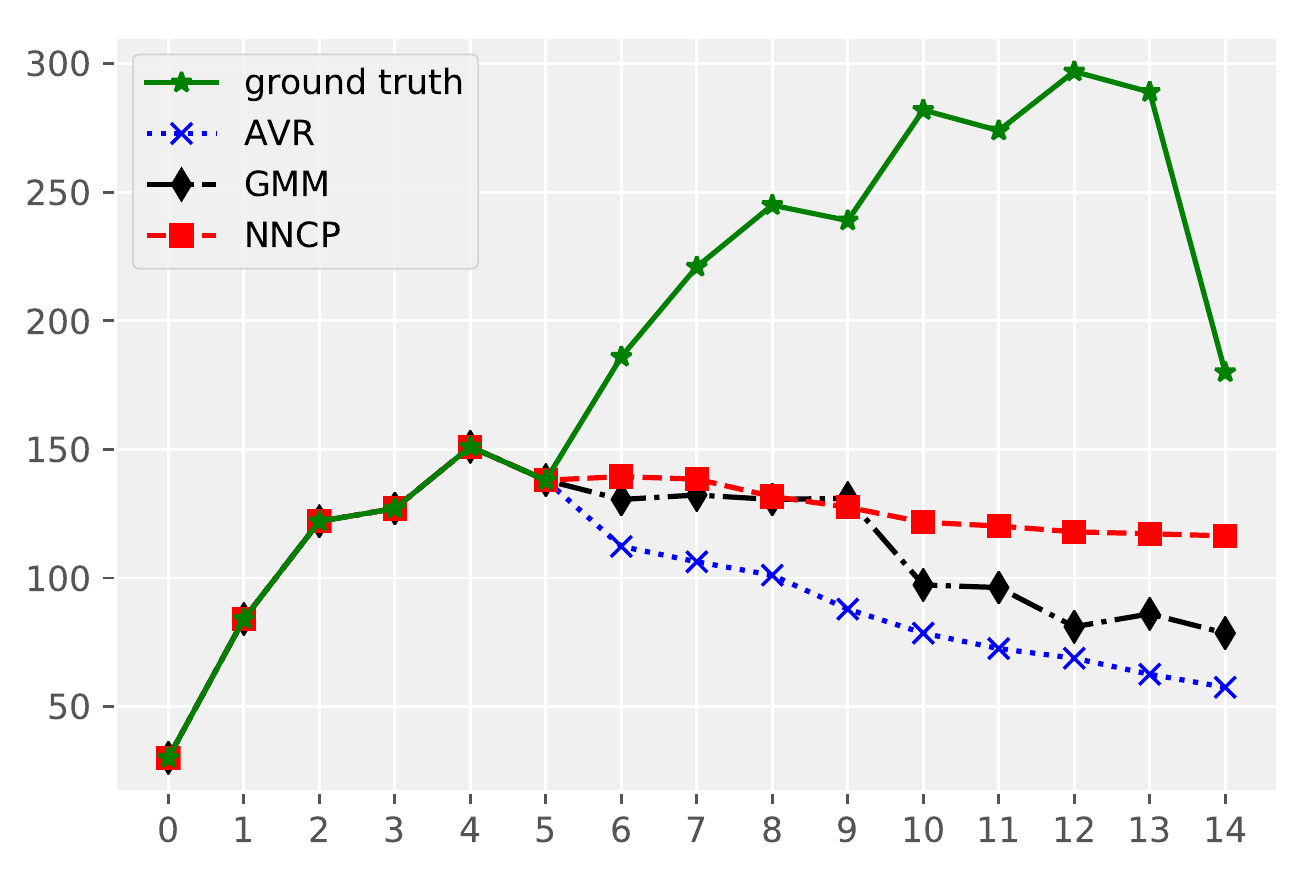}
		\caption{sample paper from NEJM}
	\end{subfigure}
	\begin{subfigure}{0.32\textwidth}
		\includegraphics[width=\linewidth]{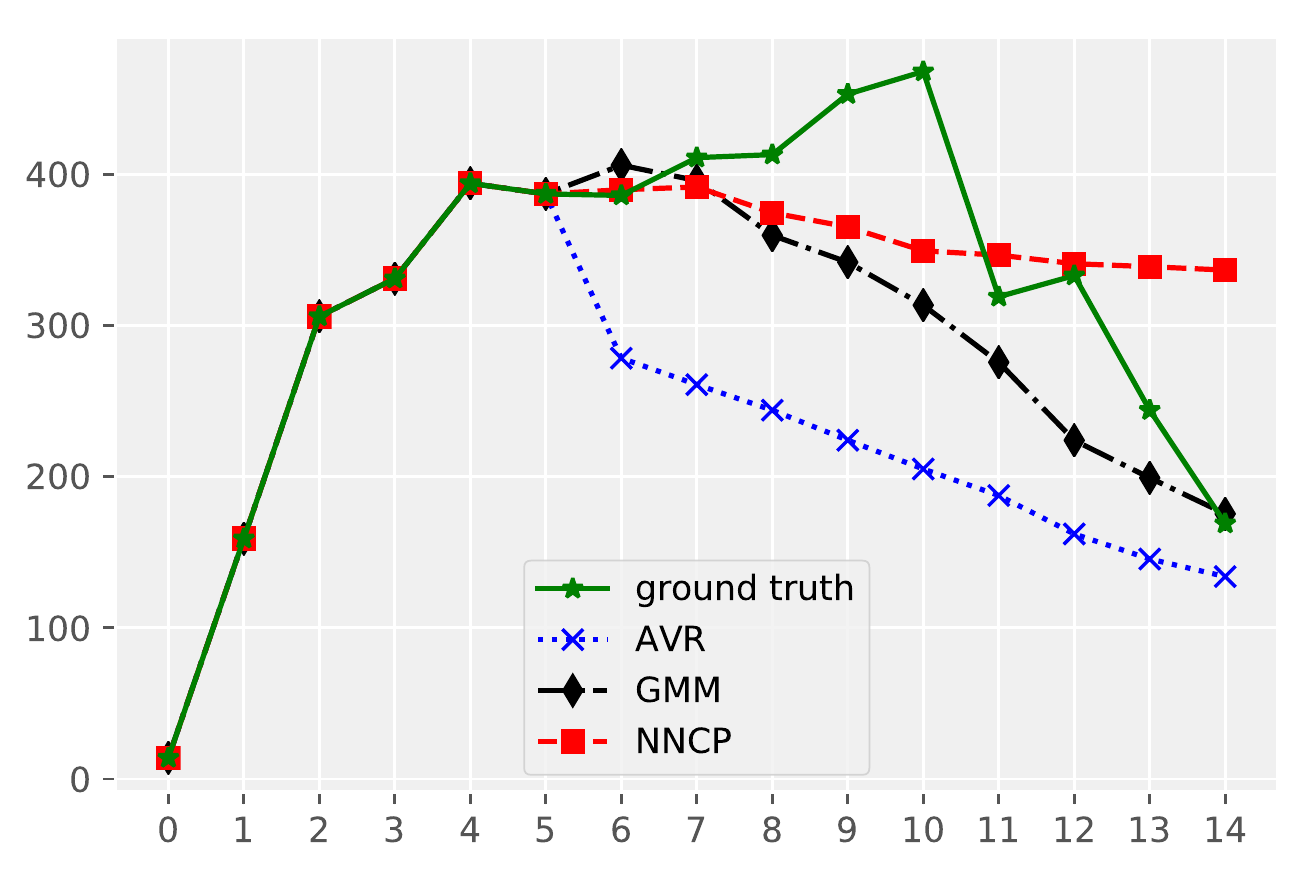}
		\caption{sample paper from NEJM}
	\end{subfigure}
	\begin{subfigure}{0.32\textwidth}
		\includegraphics[width=\linewidth]{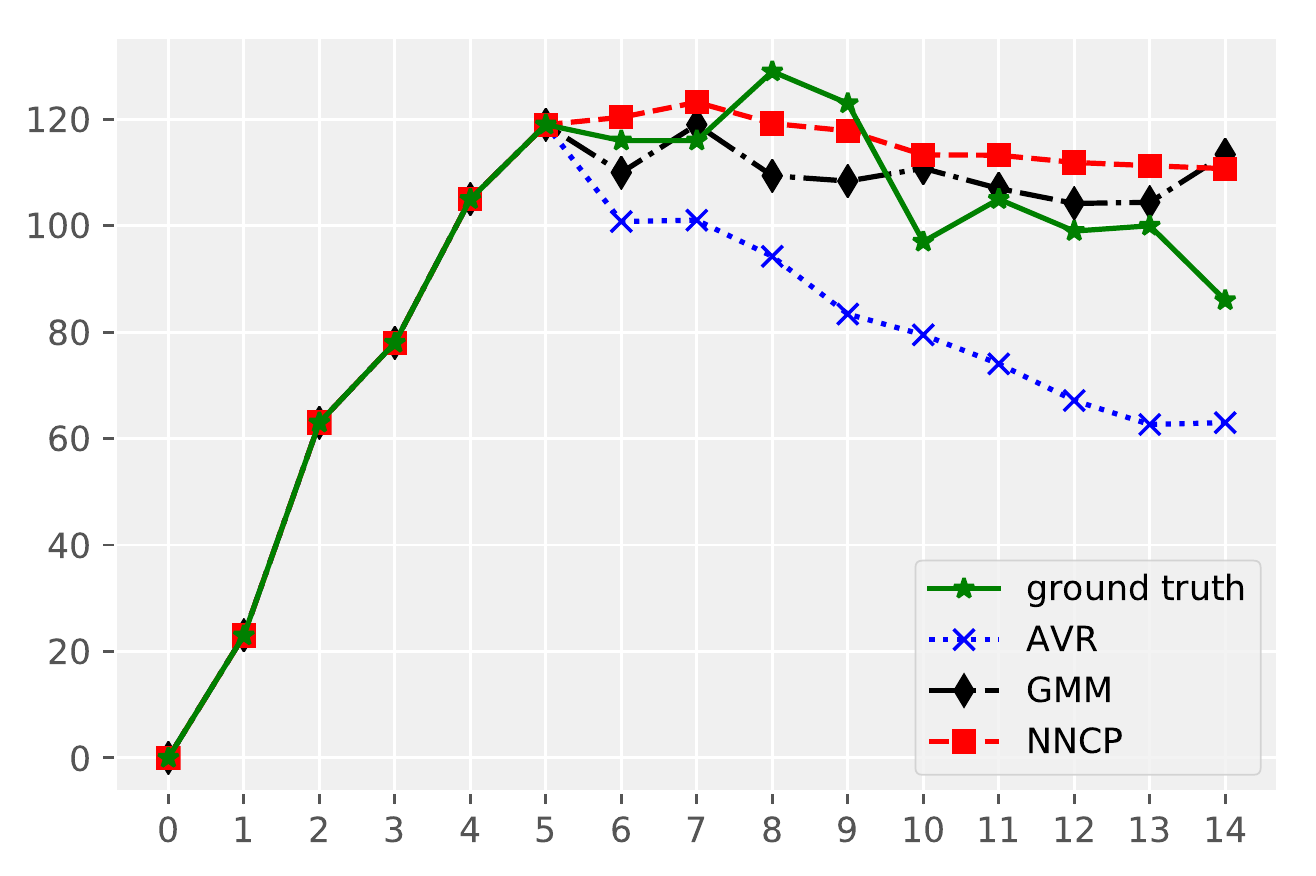}
		\caption{sample paper from NEJM}
	\end{subfigure}
	
	\begin{subfigure}{0.32\textwidth}
		\includegraphics[width=\linewidth]{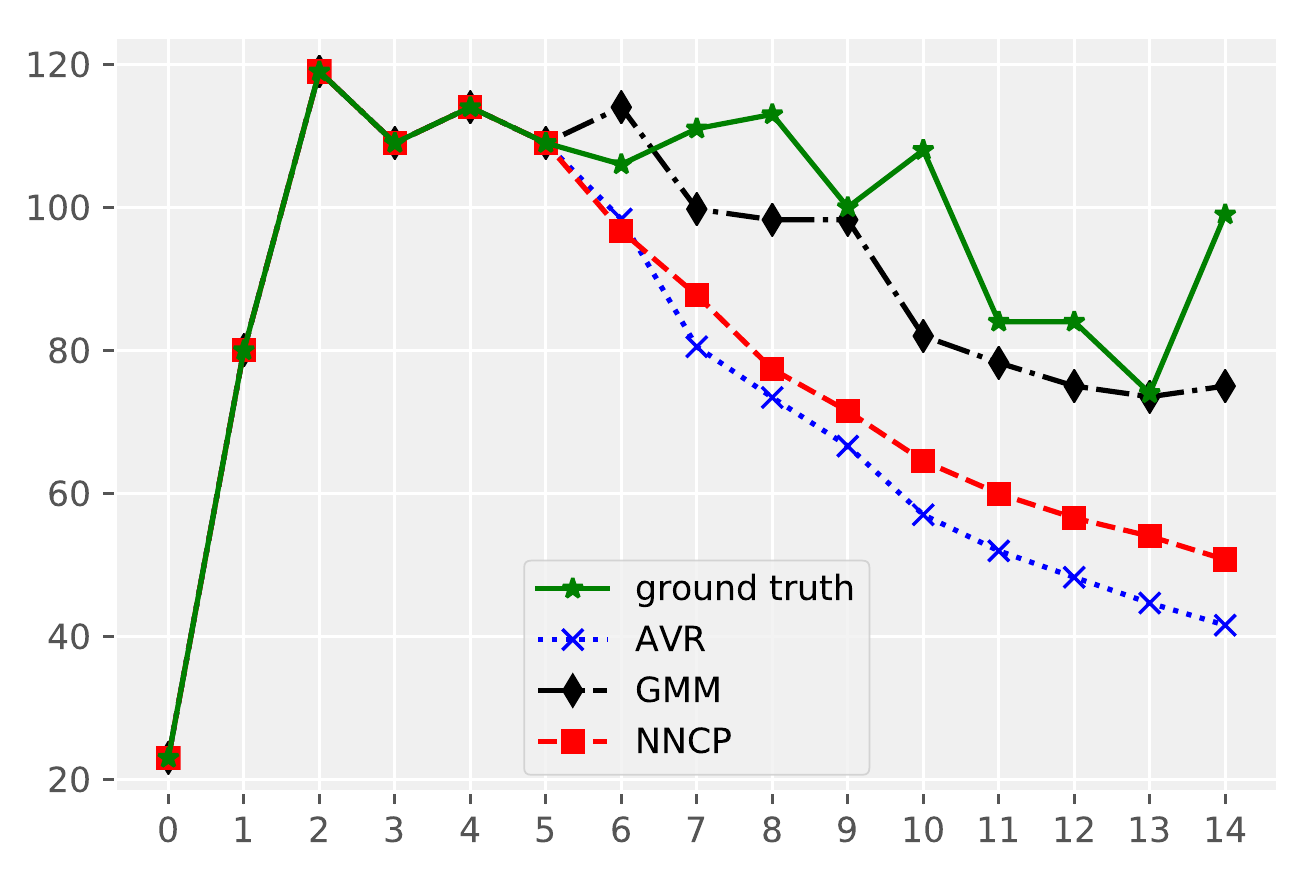}
		\caption{sample paper from Cell}
	\end{subfigure}
	\begin{subfigure}{0.32\textwidth}
		\includegraphics[width=\linewidth]{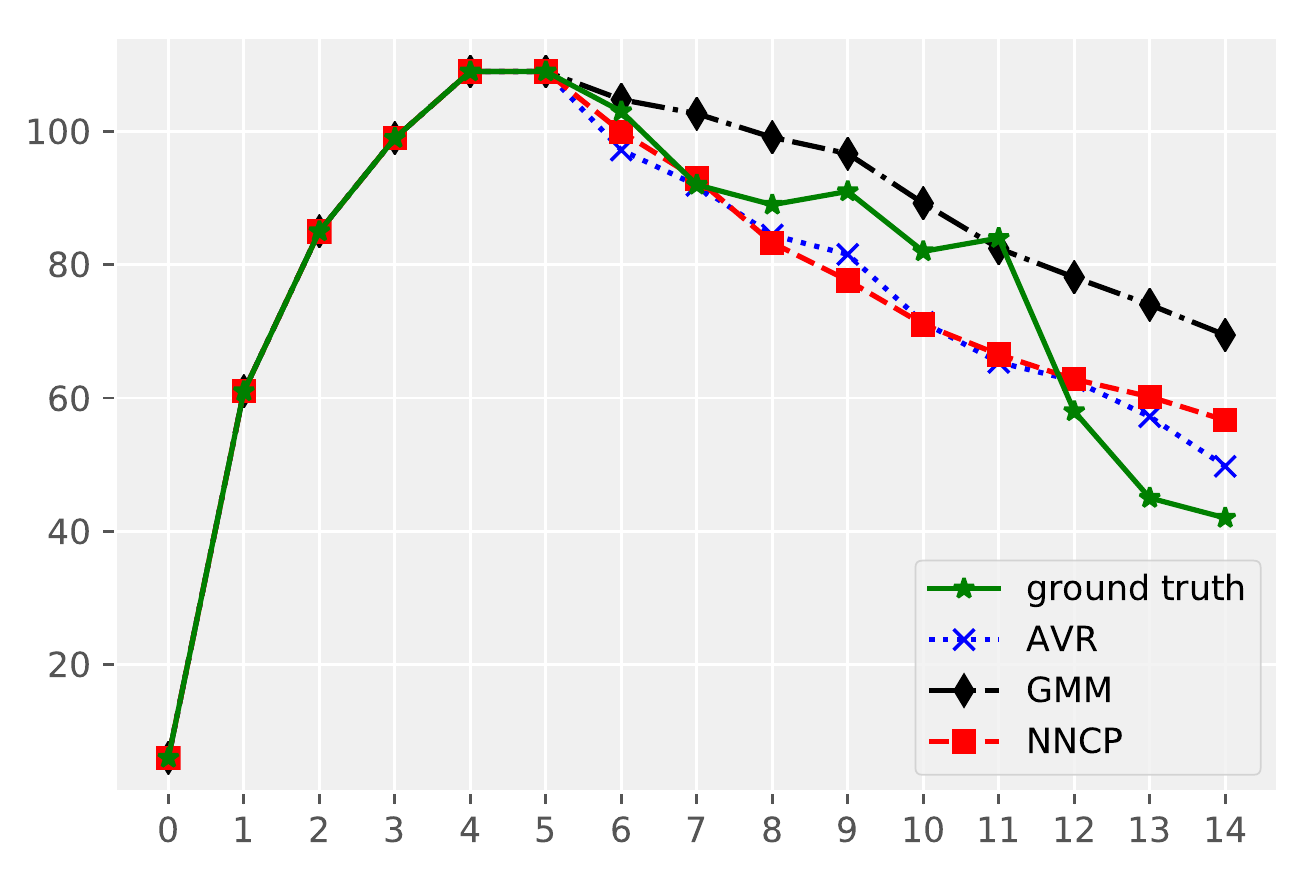}
		\caption{sample paper from Cell}
	\end{subfigure}
	\begin{subfigure}{0.32\textwidth}
		\includegraphics[width=\linewidth]{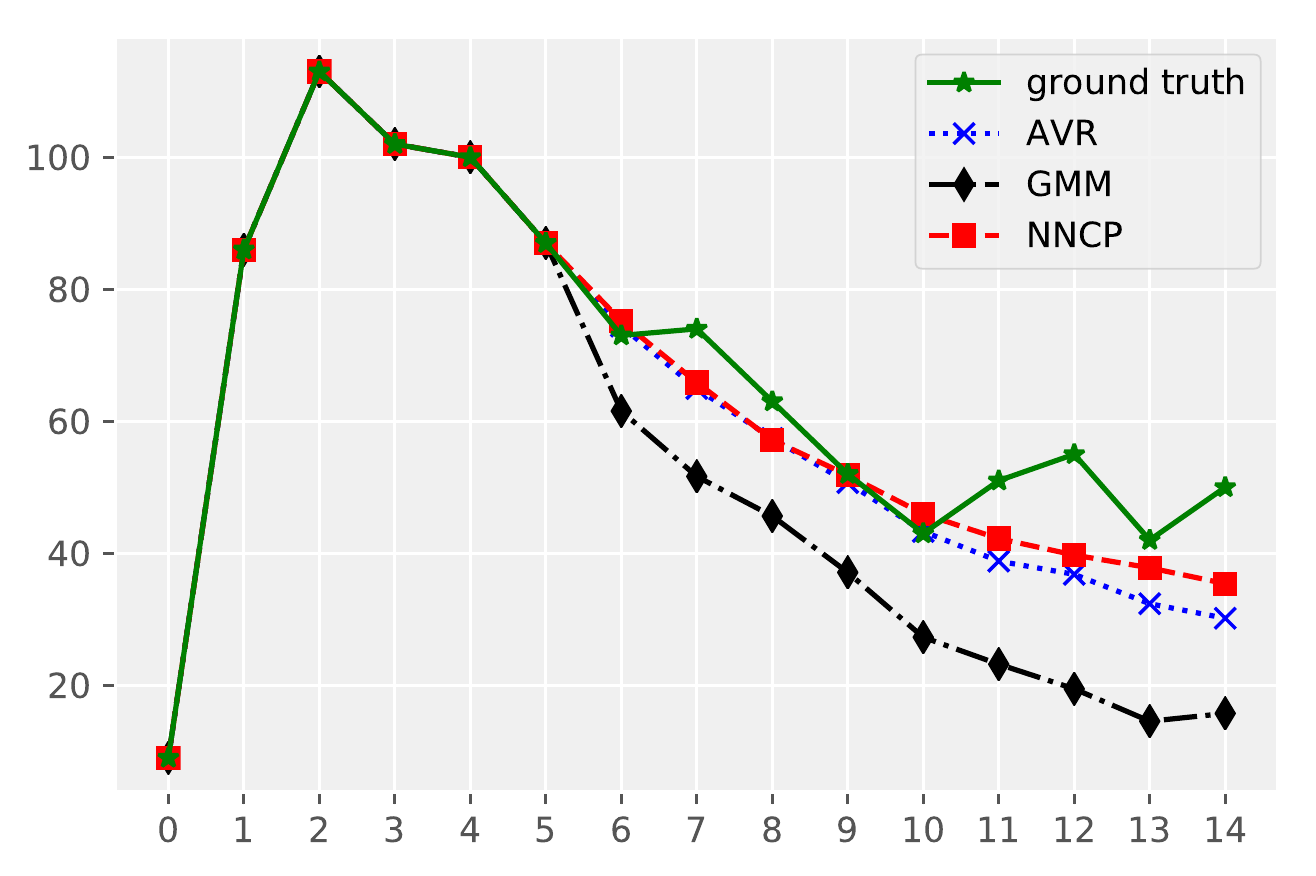}
		\caption{sample paper from Cell}
	\end{subfigure}
	
	\begin{subfigure}{0.32\textwidth}
		\includegraphics[width=\linewidth]{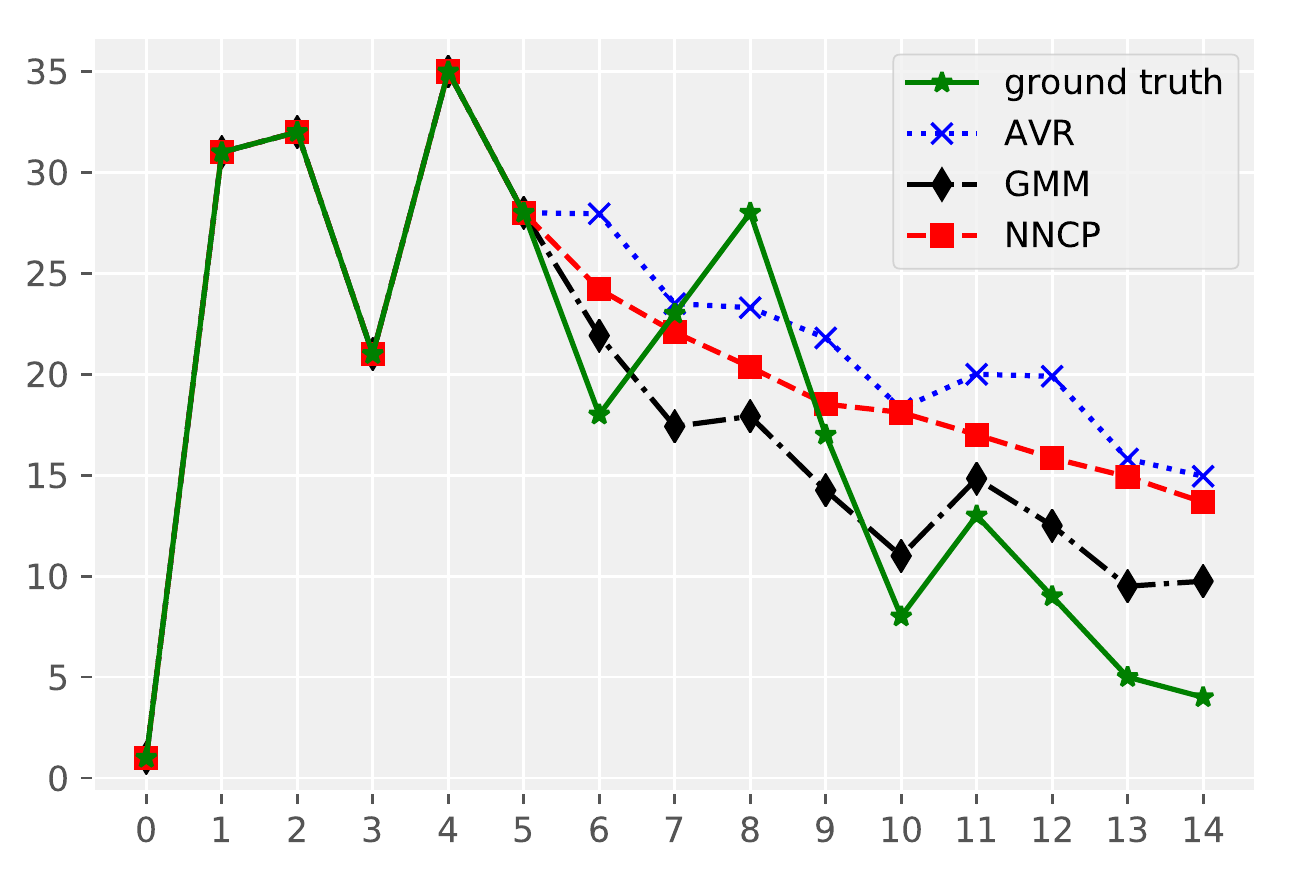}
		\caption{sample paper from PNAS}
	\end{subfigure}
	\begin{subfigure}{0.32\textwidth}
		\includegraphics[width=\linewidth]{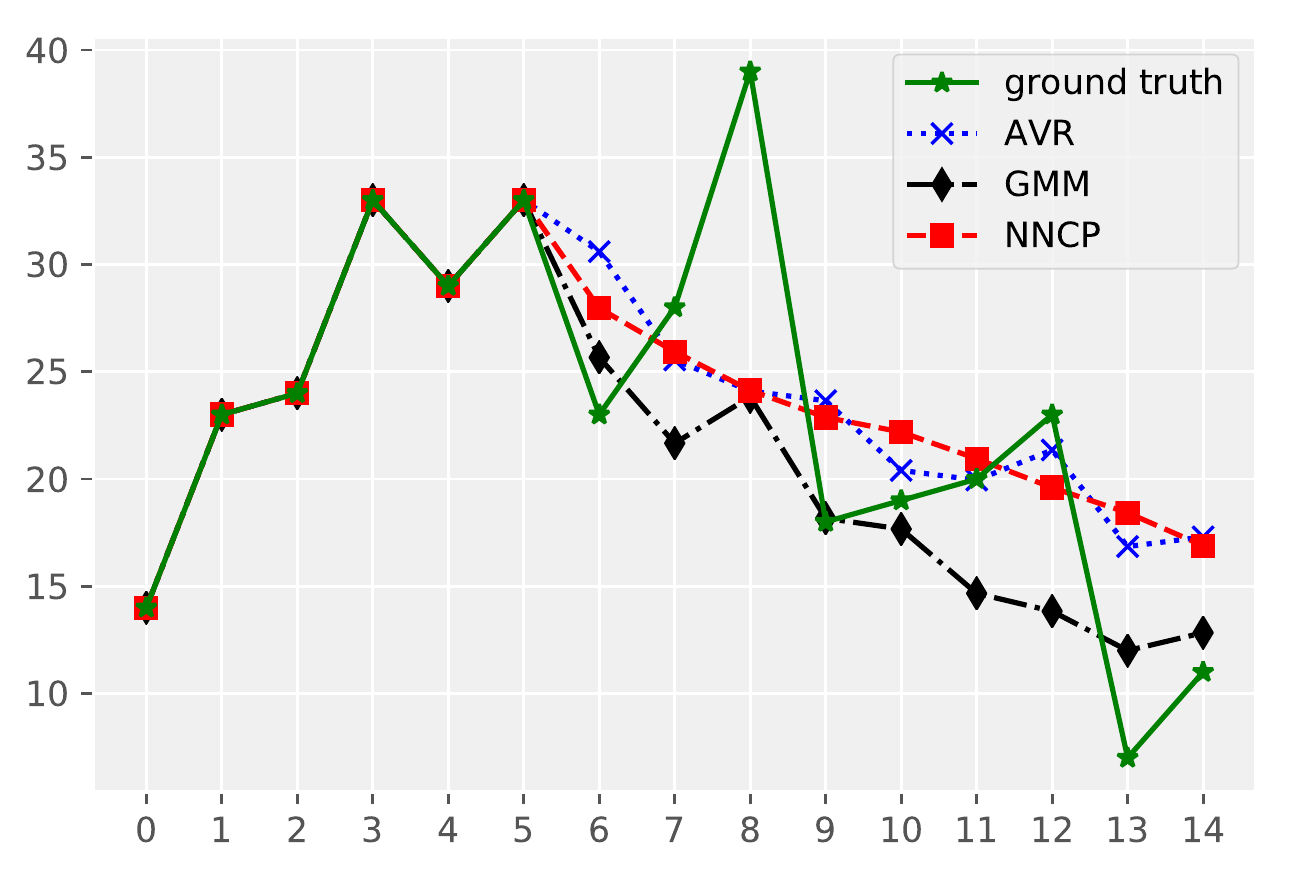}
		\caption{sample paper from PNAS}
	\end{subfigure}
	\begin{subfigure}{0.32\textwidth}
		\includegraphics[width=\linewidth]{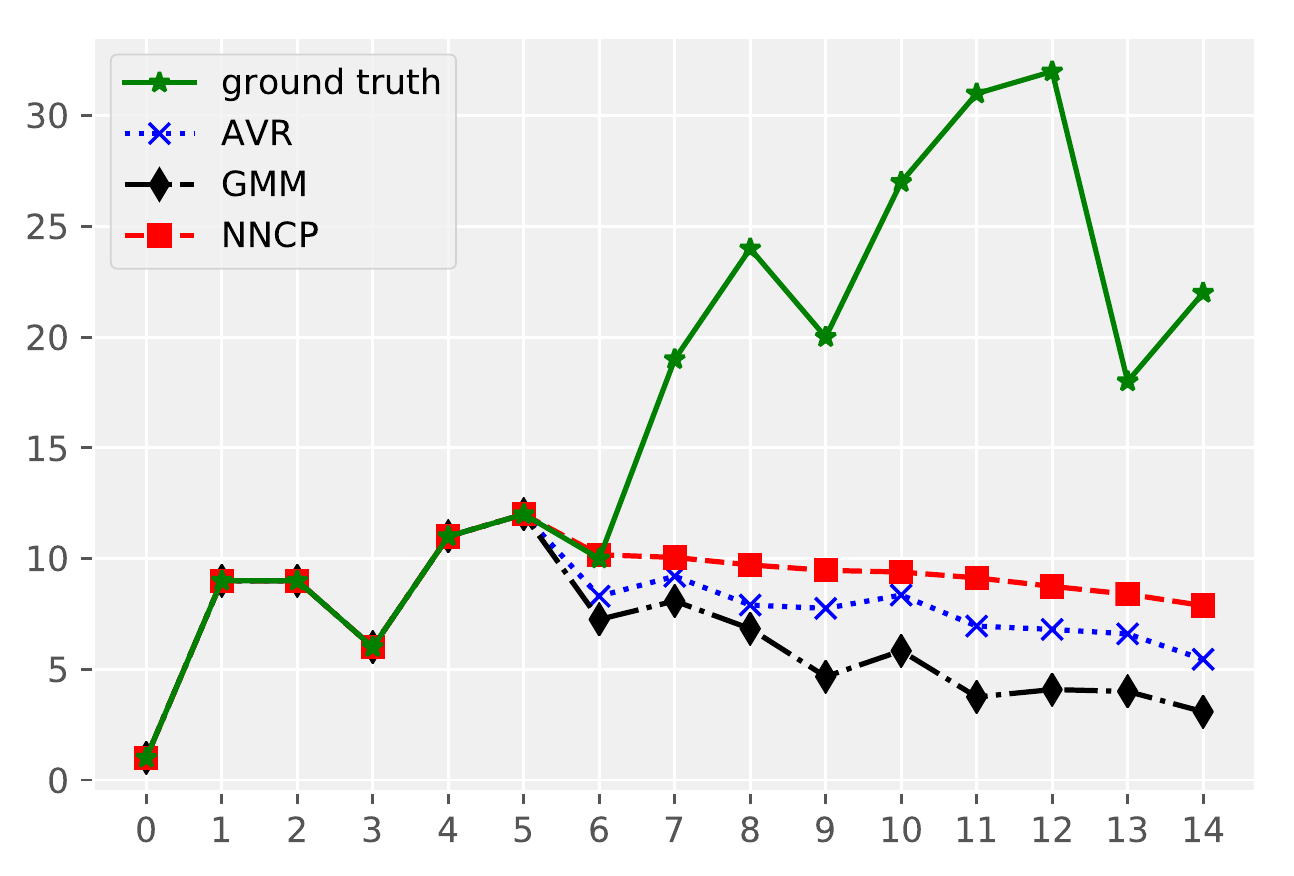}
		\caption{sample paper from PNAS}
	\end{subfigure}

	\caption{Comparison of different methods for citation count prediction of 15 randomly sampled papers.}
	\label{fig:sampledpapershistory}
	
\end{figure}

As the first motivating experiment, we employed the proposed method and the baselines in order to predict the citation count of 15 randomly selected papers among the 100 highly cited papers of our dataset. Figure \ref{fig:sampledpapershistory} illustrates the actual citation counts (the ground truth) along with the predictions of different methods when $k=5$, i.e., the citation count history up to the fifth year after publication is utilized as the input of the prediction algorithms. As the figures show, there is no simple and trivial pattern of citations in different sampled papers and therefore, predicting the future citations of a paper is not an easy task. Moreover, the proposed method outperforms the baselines (NNCP is usually the most close line to the ground-truth) for most of the sampled papers. After this motivating experiment, comprehensive and quantitative experiments are following.

\begin{figure}%
	\centering
	\begin{subfigure}[b]{0.6\textwidth}
		\includegraphics[width=\textwidth]{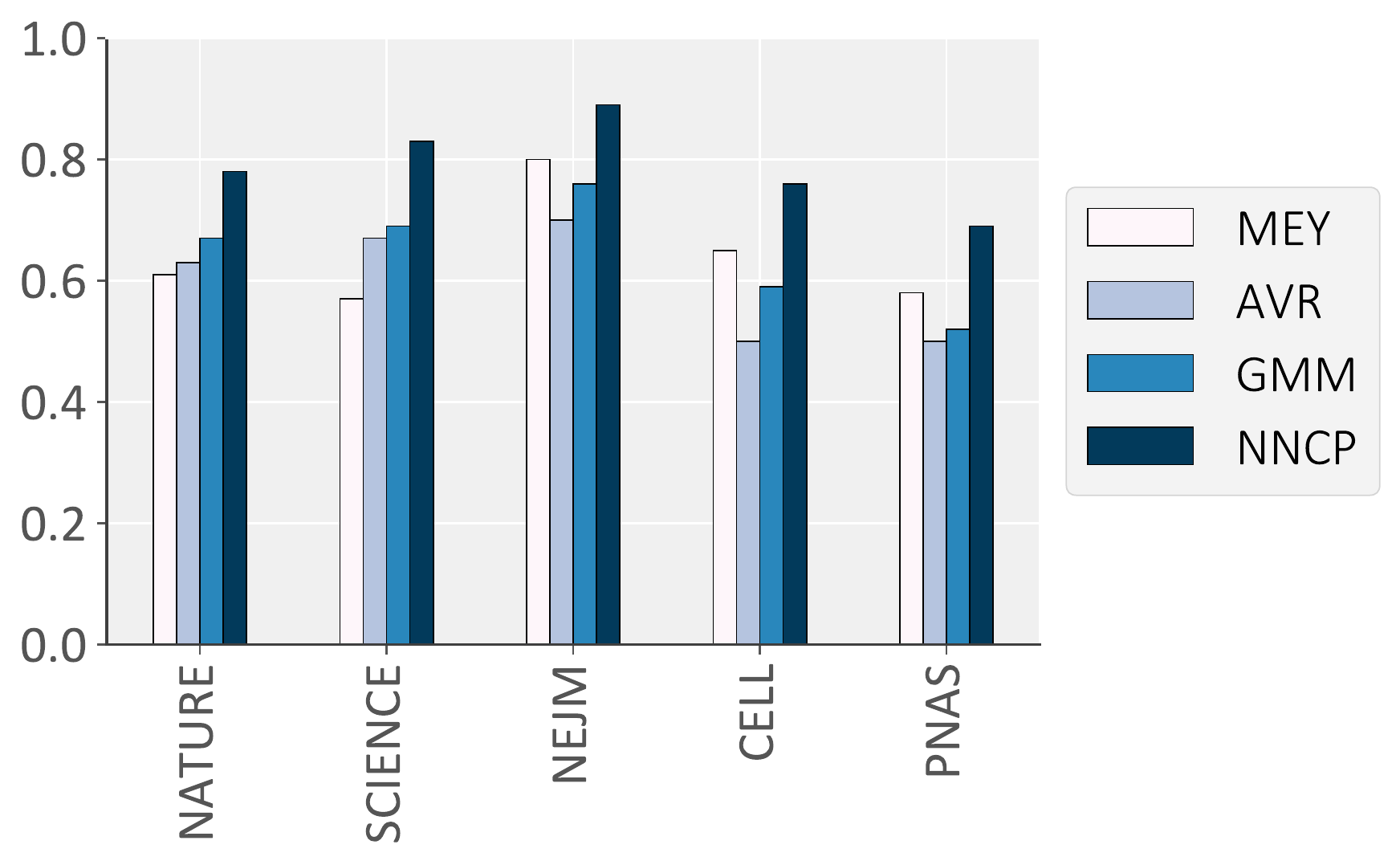}
		\caption{Average yearly $R^2$ score}
		\label{fig:5year-r2score}
	\end{subfigure}
	\begin{subfigure}[b]{0.6\textwidth}
		\includegraphics[width=\textwidth]{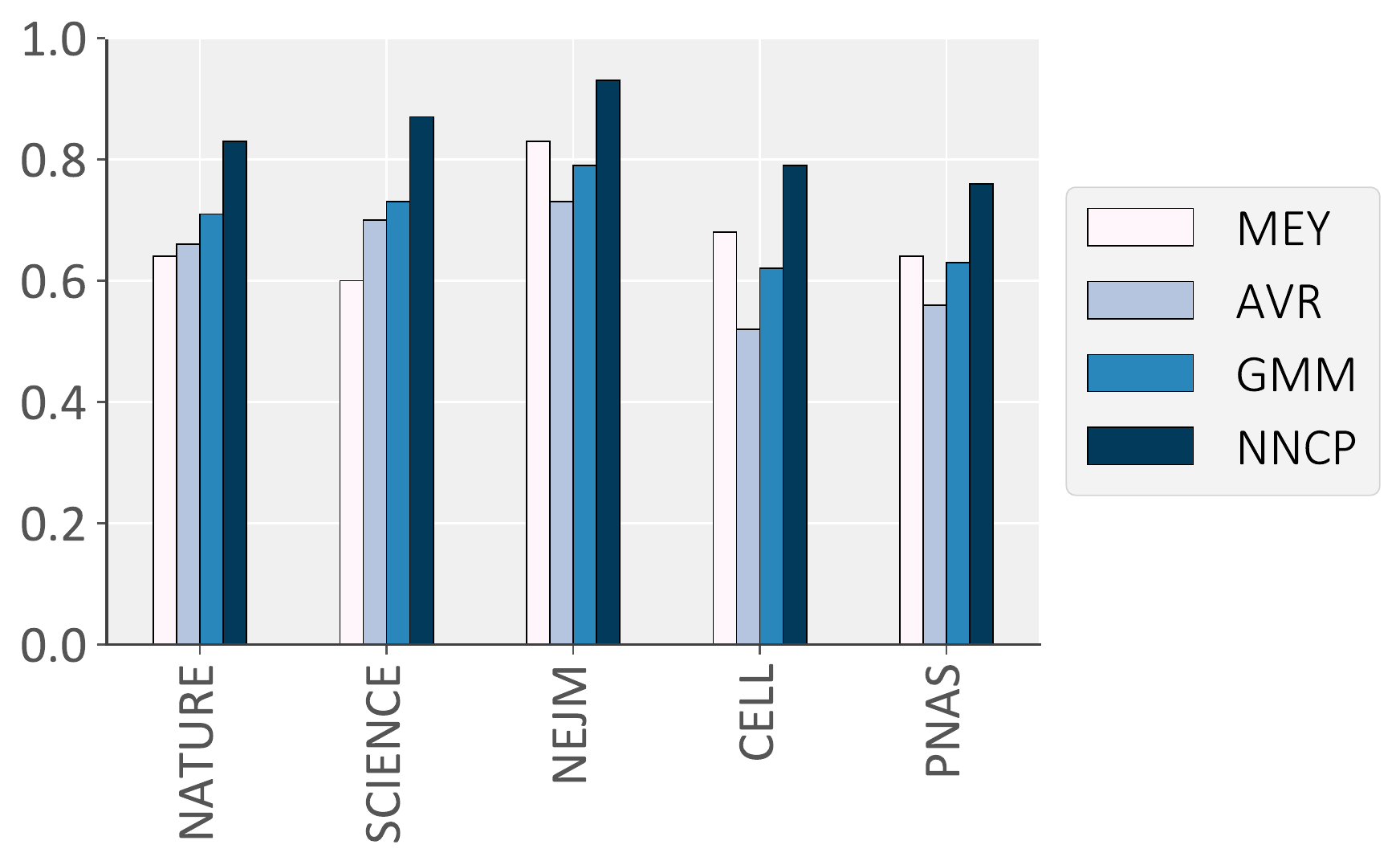}
		\caption{Total $R^2$ score}
		\label{fig:5year-r2score-total}
	\end{subfigure}
	\\
	\begin{subfigure}[b]{0.6\textwidth}
		\includegraphics[width=\textwidth]{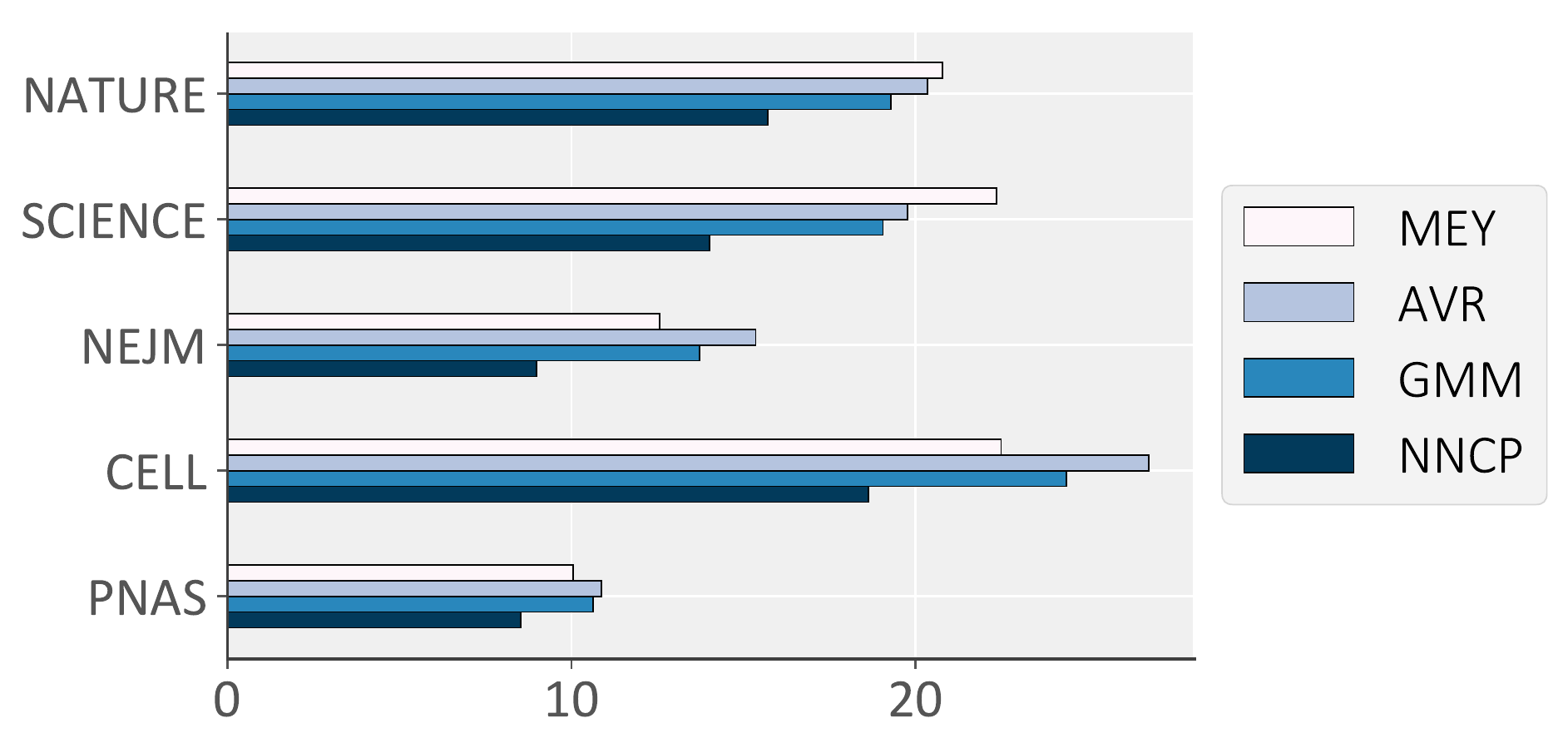}
		\caption{Average yearly RMSE}
		\label{fig:5year-rmse}
	\end{subfigure}
	\begin{subfigure}[b]{0.6\textwidth}
		\includegraphics[width=\textwidth]{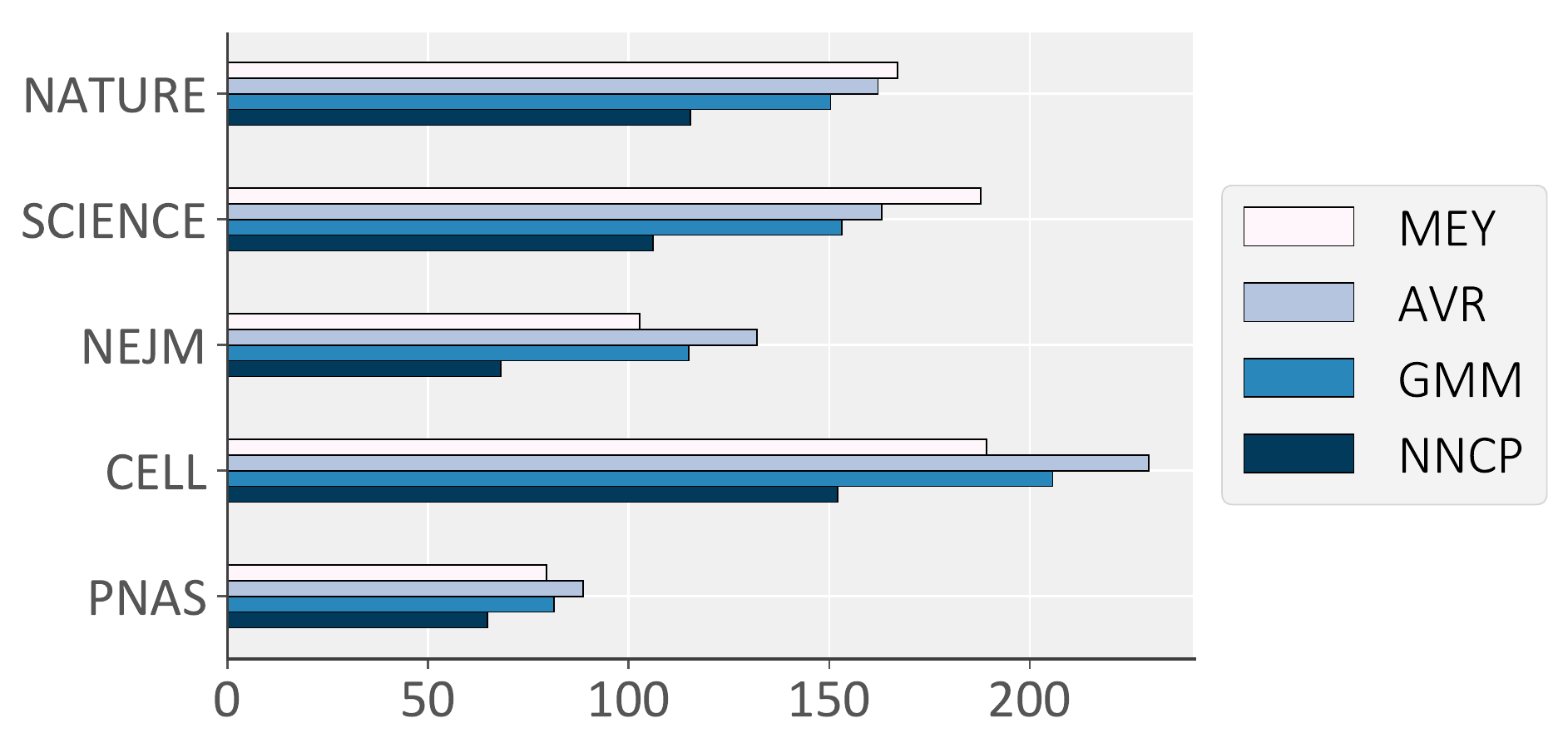}
		\caption{Total RMSE}
		\label{fig:5year-rmse-total}
	\end{subfigure}
	\caption{Evaluation results for proposed method and the baselines in $k$=5 mode.}
	\label{figure:5yearmode}
\end{figure}

In the second experiment, we set $n=14$ and $k=5$ (refer to Table \ref{table:symbols}) which means the citation counts up to the fifth year after publication are used to predict the citation count until the 14th year after publication. In this experiment, the accuracy of the methods are measured in two modes: First, the accuracy of $\hat{C}$ and second, the average accuracy of $\hat{c_i}$ values. In other words, the average accuracy of the yearly predictions is measured ($\hat{c_i}$) along with the aggregated accuracy of the predicted values ($\hat{C}$). Figure \ref{figure:5yearmode} illustrates the evaluation results of this experiment. As the figure shows, the proposed method (named NNCP) results in higher values of $R^2$ in both yearly and aggregated (total) modes, and this fact is consistent for all of the considered journals. Moreover, the proposed method results in less RMSE value for all the journals in both yearly and total modes. Therefore, the proposed method outperforms all the baseline methods in $k=5$ for all of the considered journals. 

\begin{figure}%
	\centering
	\begin{subfigure}[b]{0.6\textwidth}
		\includegraphics[width=\textwidth]{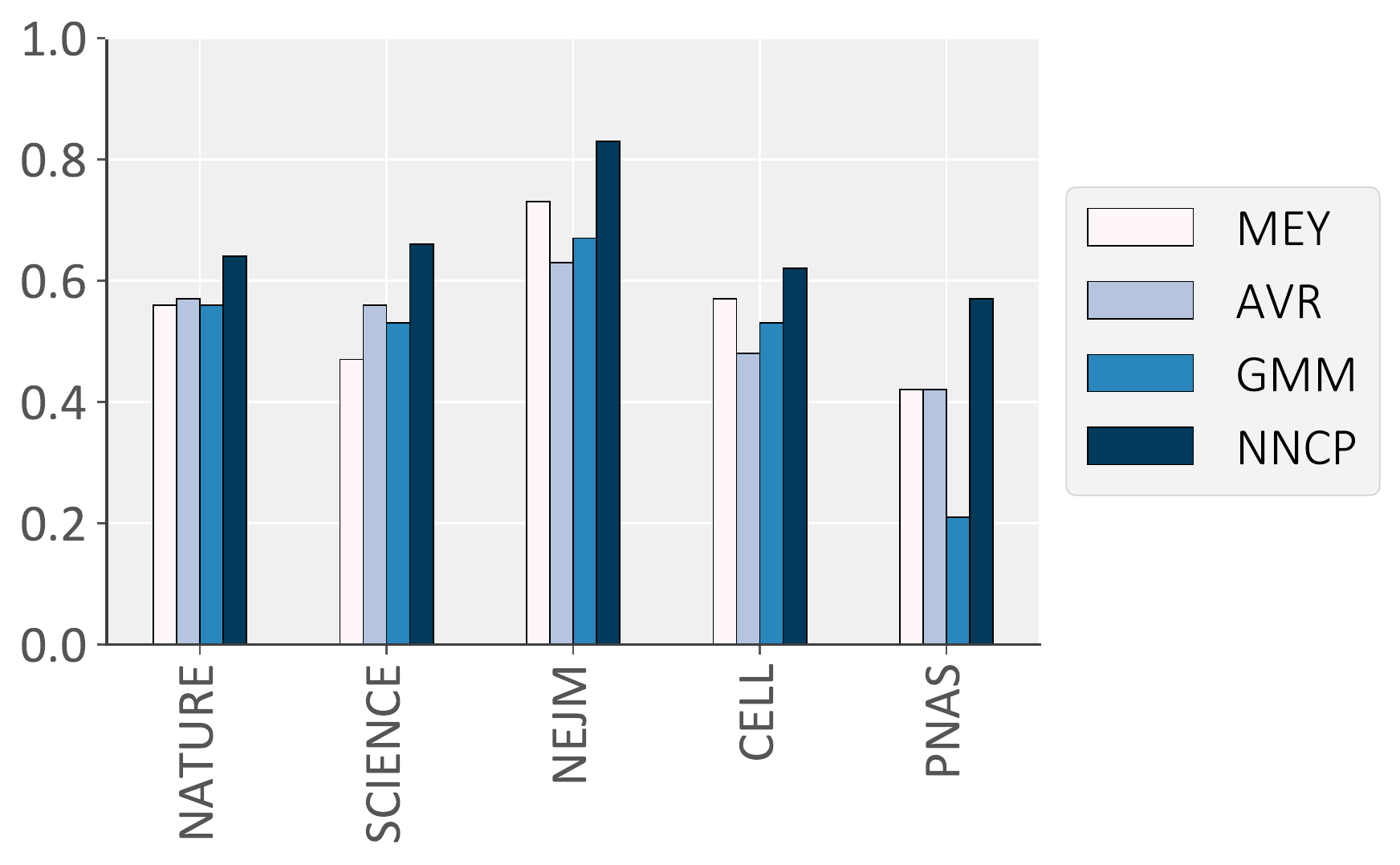}
		\caption{Average yearly $R^2$ score}
		\label{fig:3year-r2score}
	\end{subfigure}
	\begin{subfigure}[b]{0.6\textwidth}
		\includegraphics[width=\textwidth]{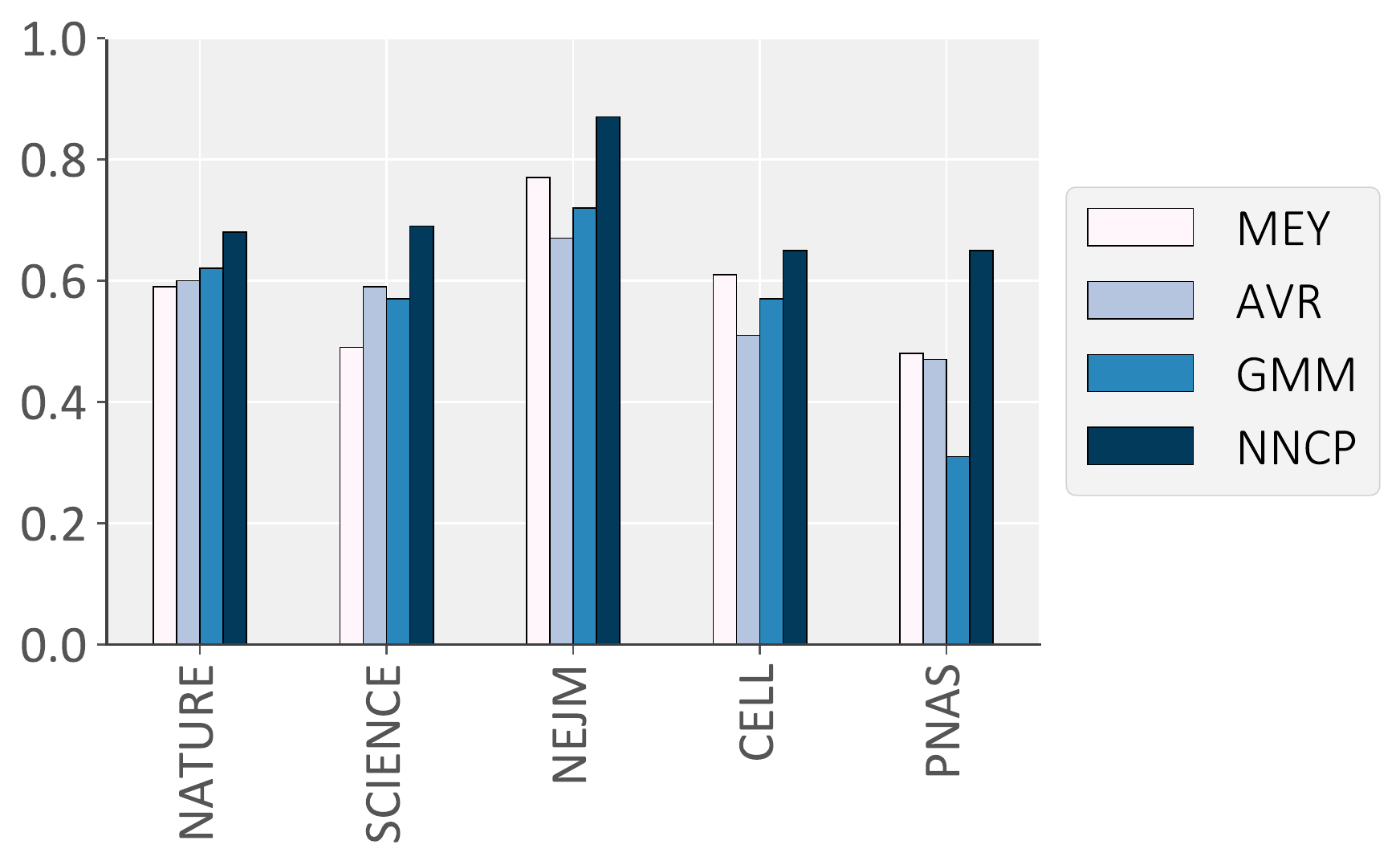}
		\caption{Total $R^2$ score}
		\label{fig:3year-r2score-total}
	\end{subfigure}
	\\
	\begin{subfigure}[b]{0.6\textwidth}
		\includegraphics[width=\textwidth]{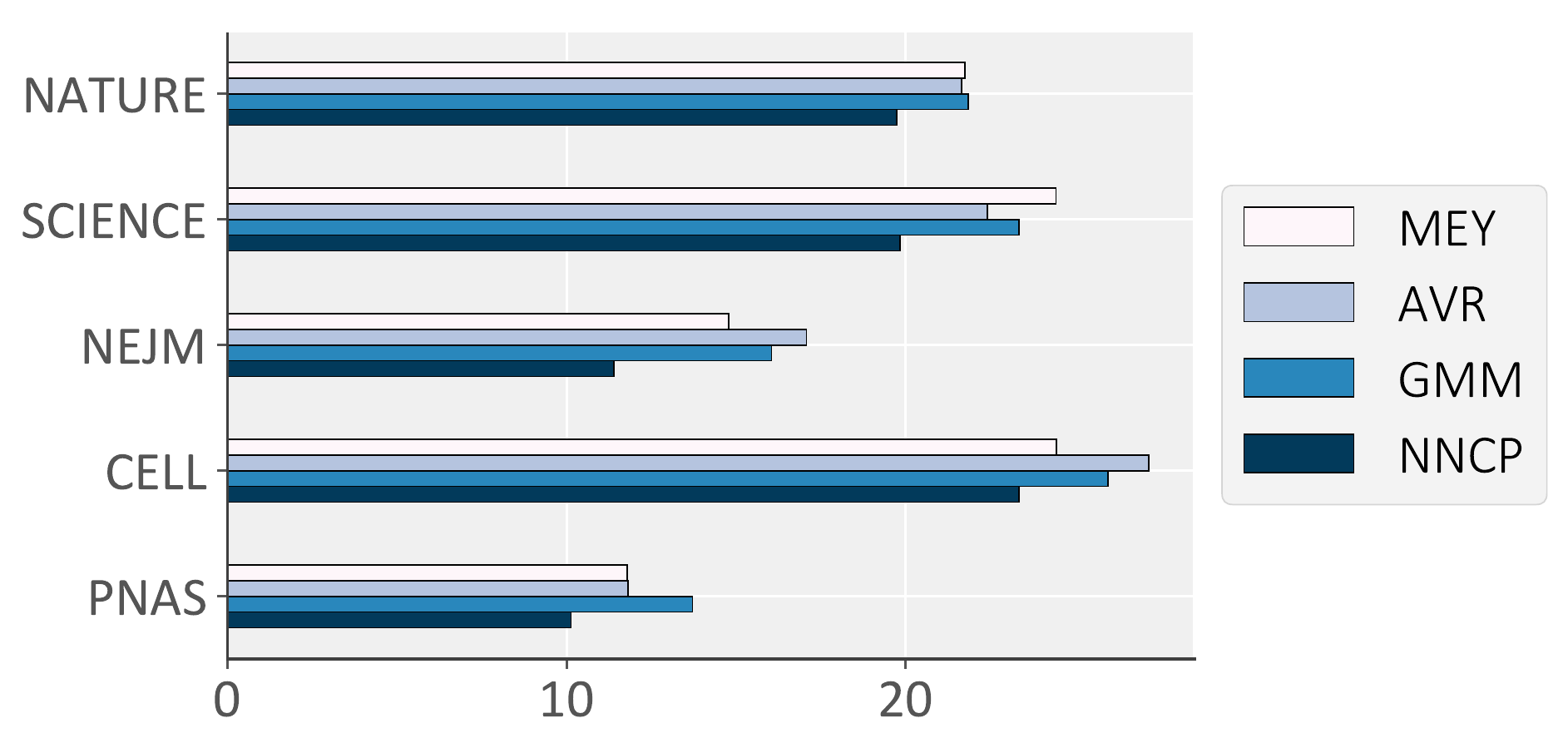}
		\caption{Average yearly RMSE}
		\label{fig:3year-rmse}
	\end{subfigure}
	\begin{subfigure}[b]{0.6\textwidth}
		\includegraphics[width=\textwidth]{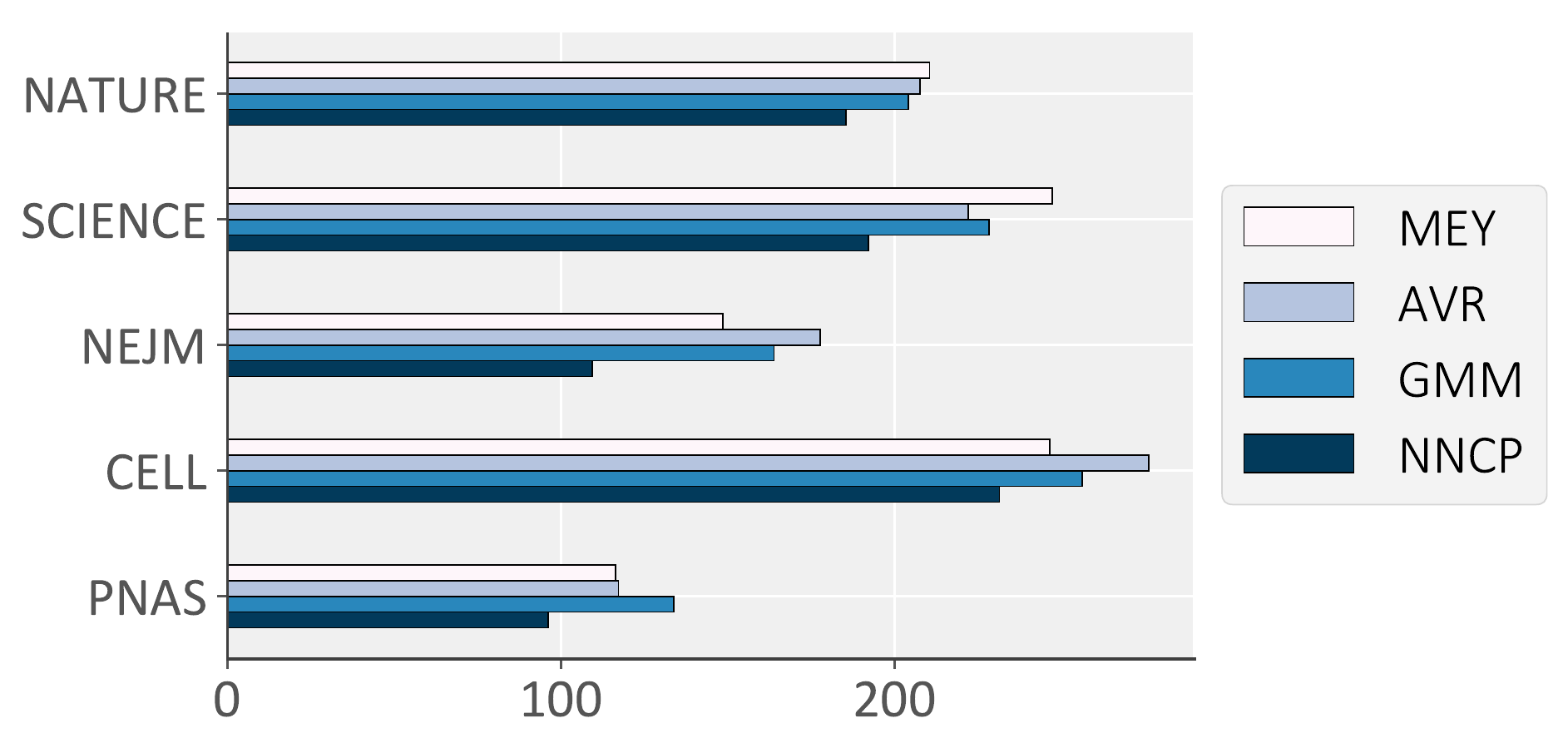}
		\caption{Total RMSE}
		\label{fig:3year-rmse-total}
	\end{subfigure}
	\caption{Evaluation results for proposed method and the baselines in $k$=3 mode.}
	\label{figure:3yearmode}
\end{figure}

In the third experiment, we limit the input information to the citation count up to the third year of publication, and we repeat the second experiment but for $k=3$. In this experiment, only the citation history up to the third year after publication is used in order to predict the citation counts in the rest of the subsequent years until the 14th year of publication. Figure \ref{figure:3yearmode} shows the evaluation results of this experiment. As the figure shows, the proposed method still outperforms all the baselines in the $k=3$ mode according to both RMSE and $R^2$ criteria for all journals of the dataset. It is also worth noting that the prediction results of all the methods are more accurate in the $k=5$ than those of the $k=3$ mode according to both RMSE and $R^2$ scores, because with $k=5$ more input information is available for the methods (citation counts of five years after publication, instead of only three years citation history) in order to investigate the citation counts in the future.

\begin{figure}%
	\centering
	\includegraphics[scale=0.45]{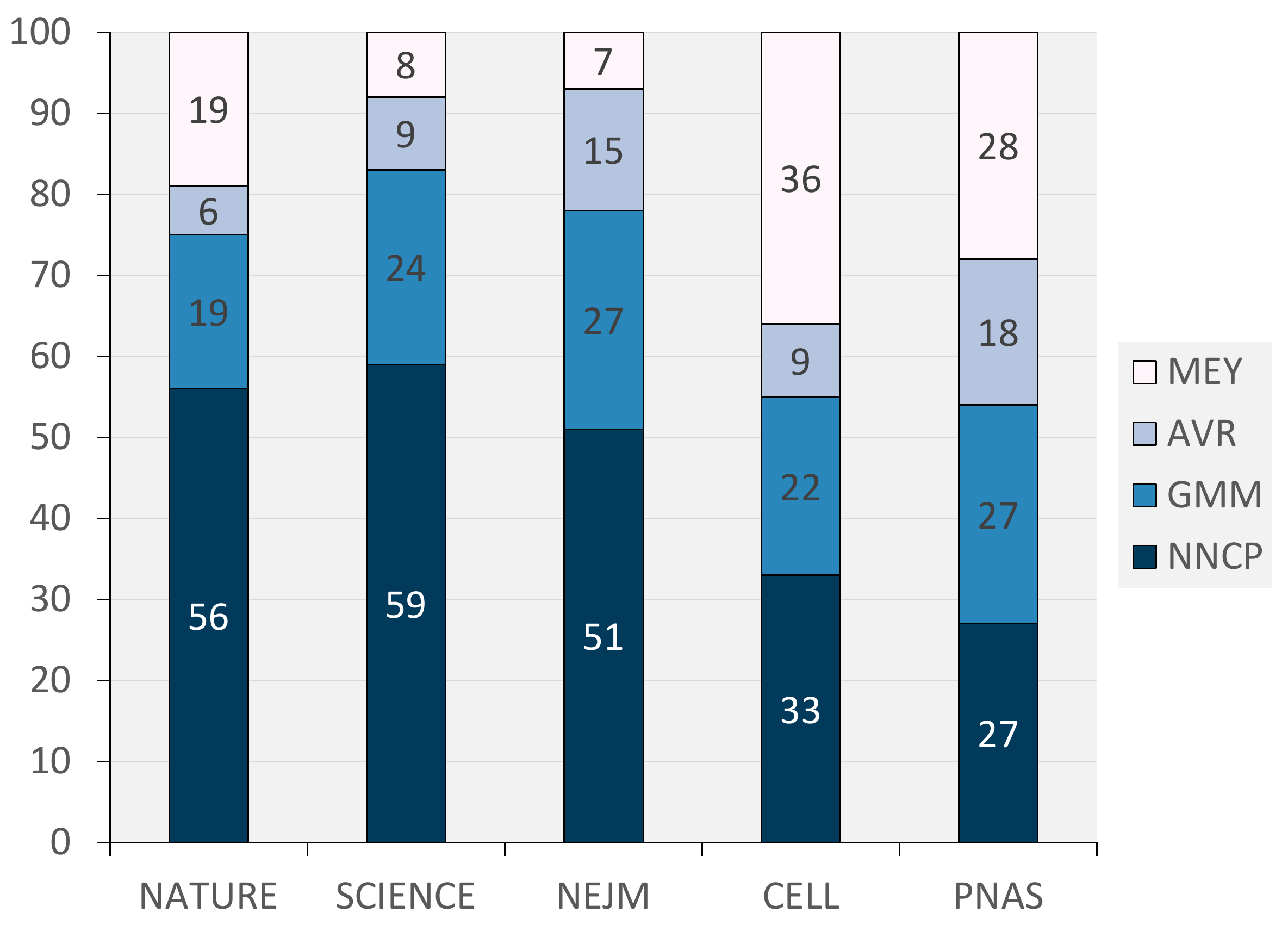}
	\caption{Frequency of best prediction score for the 100 highly cited papers when $k=5$.}
	\label{figure:top100-5years}
\end{figure}

\begin{figure}%
	\centering
	\includegraphics[scale=0.45]{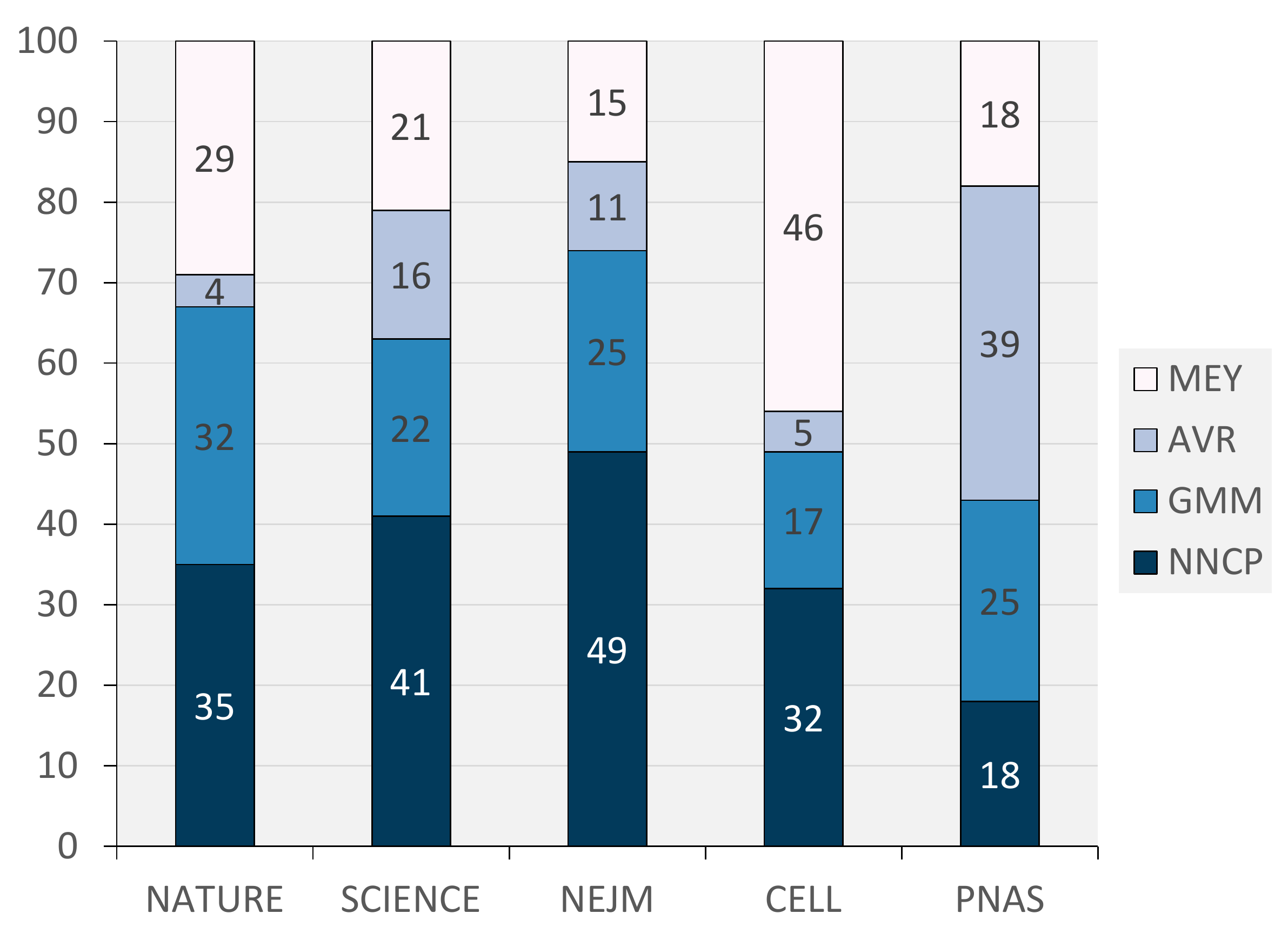}
	\caption{Frequency of best prediction score for the 100 highly cited papers when $k=3$.}
	\label{figure:top100-3years}
\end{figure}

One of the important applications of citation count prediction methods is to detect highly cited papers in advance. Some efforts exist in the literature for early detection of highly cited papers \citep{0295-5075-105-2-28002, Sarigol2014}. Therefore, in the next experiment, we evaluate the ability of the proposed method and the baselines regarding their accuracy in predicting citations of highly cited papers. We considered top 100 highly cited papers of each journal in our dataset published between 1998 and 2002 according to their total citations up to 2018. Then, we utilized the proposed method and the baselines in order to predict the citation count of those highly cited papers. We ran the experiment in two phases of $k=3$ and $k=5$, and we computed the error of the methods according to the RMSE criterion. Finally, we ranked different methods according to their accuracy of citation prediction. Figures \ref{figure:top100-5years} and \ref{figure:top100-3years} show the fraction of samples (highly cited papers) that each method has resulted the best prediction for $k=5$ and $k=3$ respectively. For example, the first bar of the Figure \ref{figure:top100-5years} shows that the proposed method (NNCP) is much more accurate than the baselines when they are employed to predict the citation count of the 100 highly cited papers of Nature. In other words, this bar shows that our proposed method results in the best citation count prediction accuracy among the baselines (in 56 out of 100 samples) for the highly cited papers of Nature. Figure \ref{figure:top100-3years} shows the result of a similar experiment but for $k=3$ (when only the citation count up to the third year after publication is utilized as the input). As Figures \ref{figure:top100-5years} and \ref{figure:top100-3years} show, in most cases, the proposed method outperforms the baseline methods in predicting the citation count of the highly cited papers, and this fact is valid in both $k=5$ and $k=3$ cases. But in the case of PNAS and Cell journals, our proposed method is not ranked first among the baselines. This is mainly because Cell and PNAS include fewer samples in our dataset (fewer number of papers). In other words, while Figures \ref{figure:5yearmode} and \ref{figure:3yearmode} showed that our proposed method outperforms all the baselines when we consider all the papers, Figures \ref{figure:top100-5years} and \ref{figure:top100-3years} show that in the case of highly-cited papers, our proposed method is more effective than the baselines when we have enough data in the experiments (e.g, in the case of Nature, Science, and NEJM, but not for Cell and PNAS). This fact is also consistent with the results of the sampled papers illustrated in Figure \ref{fig:sampledpapershistory}.

\subsection{Sensitivity Analysis}
\begin{figure}
	\centering
	\begin{subfigure}{0.80\textwidth}
		\includegraphics[width=\linewidth]{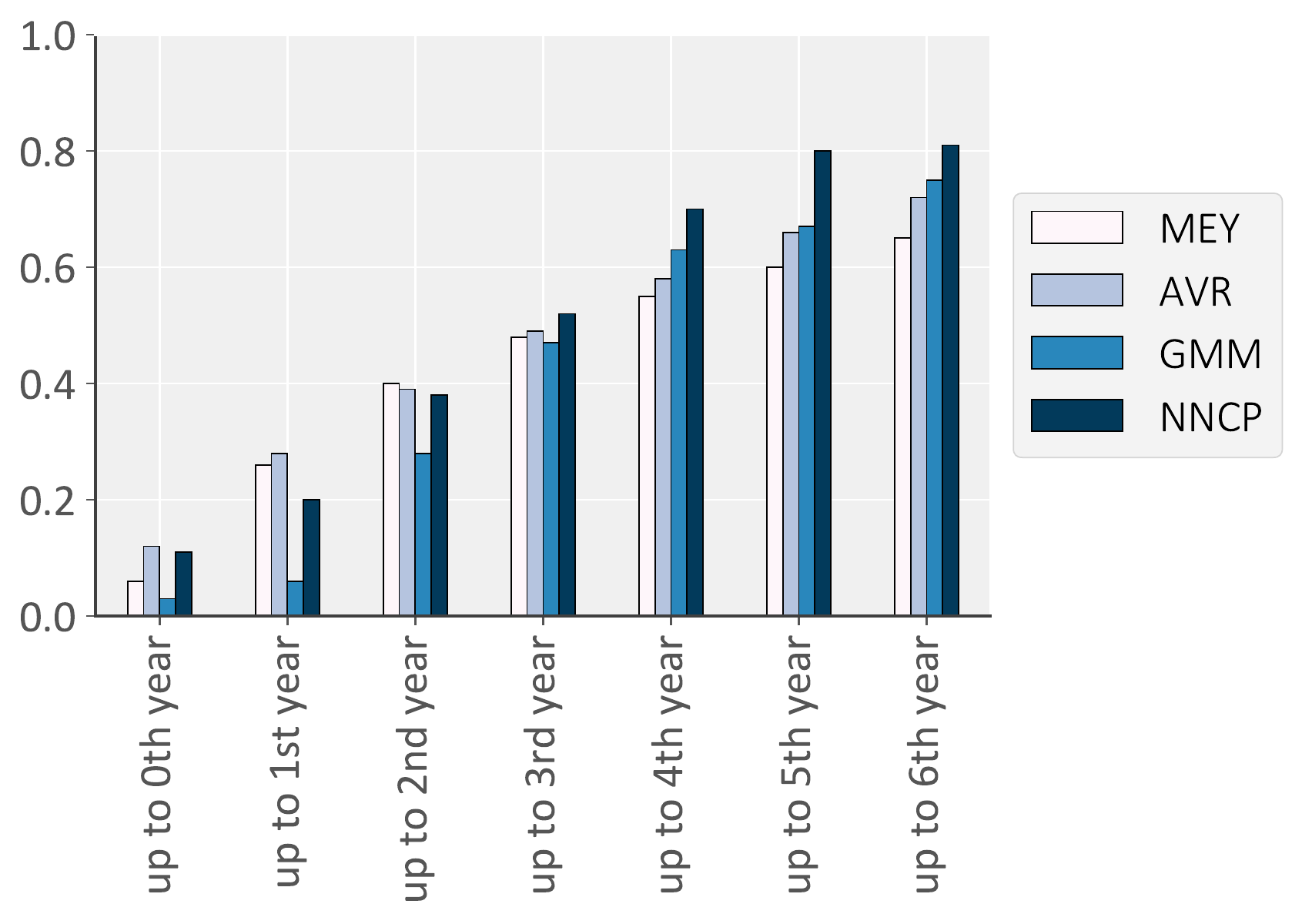}
		\caption{$R^2$ score}
	\end{subfigure}

	\begin{subfigure}{0.95\textwidth}
		\includegraphics[width=\linewidth]{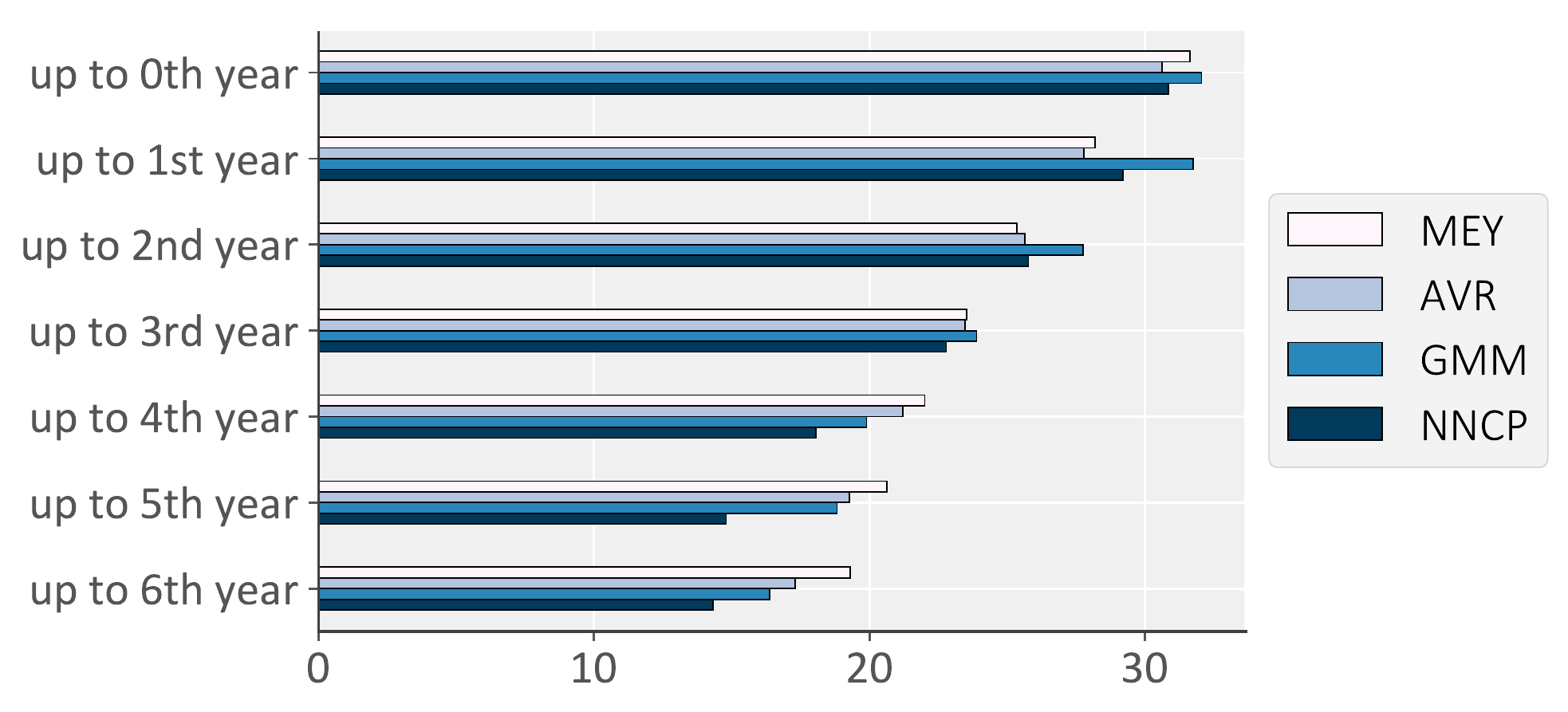}
		\caption{RMSE}
	\end{subfigure}

	\caption{Sensitivity analysis of different methods to parameter $k$. The figure shows the prediction score of different methods for predicting citations of 7th to 14th years of publication, when only the citation count of the first $k+1$ years are available, for $0 \leq k \leq 7$.}
	\label{figure:sensitivity}
\end{figure}

As the final experiment, we analyzed the overall accuracy of the considered methods according to different values of $k$ ($0 \leq k \leq 7$). In this experiment, all the papers of the dataset published in the five journals within 1980 to 1997 are considered as the training set, and all the papers of the dataset published from 1998 to 2002 are used as the test set. Therefore, this experiment also includes the largest training set and test set utilized in this research. In this evaluation scenario, different methods are utilized in order to predict the citation count of the papers in the test set from the 7th to 14th year after the publication. In other words, the methods are employed to predict $c_7, c_8, ... ,c_{14}$ values. Figure \ref{figure:sensitivity} shows the overall accuracy of different methods according to the $R^2$ score and RMSE with respect to different values of $k$. When $k=0$ the citation count of the papers only in the publication year is utilized as the input of the methods, and when $k=6$, citation count history of the publication year until the 6 year after publication is utilized. As the figure shows, by increasing $k$, more information about the citation history of the papers in the early years of publication is fed to the methods and therefore, the accuracy of different methods improves, as expected. Moreover, the proposed method can outperform baselines when the citation history of the papers is available at least for the first three years (when $k \geq 3$). Therefore, when sufficient information about citation history of the papers is obtainable, the proposed method outperforms the baselines.

\subsection{Discussion}
We reviewed several experiments which show effectiveness of the proposed method when it compared with the baseline methods. Among the baselines, MEY is a na\"{\i}ve prediction method, but its accuracy is considerable in some limited situations. For example, when we have a small dataset with limited number of sample data and we want to predict the citation count of highly cited papers, MEY method is effective in comparison to other examined methods (see the case of Cell and PNAS journals in Figures \ref{figure:top100-5years} and \ref{figure:top100-3years}). MEY baseline helps investigate situations in which basic heuristics are effective in citation prediction and moreover, it serves as a simple baseline to evaluate other more complicated methods. 

AVR and GMM baselines, follow a local modeling technique called lazy learning in which no model is actually learned from the training data, but the training data are memorized instead. One disadvantage of lazy learning techniques, such as the cases of AVR and GMM, is the need to a large space for storing the entire training dataset. Moreover, noisy training data may decrease the accuracy of such methods. Additionally, AVR and GMM methods actually utilize the citation pattern of only a few (similar) existing papers in order to predict the citation count of the target paper, while NNCP makes use of the whole dataset of papers in order to learn the citation pattern of the scientific papers. It is also worth noting that lazy learning methods are usually slower than their eager learning counterparts. AVR and GMM methods have to compare the input data (the citation pattern of the target paper in its early publication years) with many samples of the dataset in order to find the most similar papers, and this process is repeated for each new query (new target paper). On the other hand, our proposed method once learns a prediction model from the training data and then, applying the model for a new data sample (a new target paper) will be an easy and fast task. In other words, NNCP is considerably faster than AVR and GMM methods in predicting the citation counts, because NNCP has passed a training phase and it has learned a prediction model which is needless to re-examine a large set of training data. Therefore NNCP outperforms AVR and GMM with respect to both efficiency and accuracy.

\section{Conclusion}
\label{sec:conclusion}
In this paper, we proposed a novel method for citation count prediction, which is based on artificial neural networks. We employed modern deep learning techniques (such as RNNs and sequence-to-sequence model) in order to learn a prediction method based on the sequence pattern of the citations from early years of publication of a paper. The comprehensive evaluations show that the proposed method outperforms state-of-the-art methods of citation count prediction with respect to the accuracy of the prediction and the ability to predict the citations of highly-cited papers. Many challenges arise when applying deep learning to the problem of citation prediction. Those challenges include: Mapping the citation problem into a deep learning problem, designing the learning and experimentation scenarios for this specific problem, choosing the evaluation criteria, selecting the best fitting deep learning architecture for this specific application, and tuning the hyperparameters of the selected neural network architecture.

As the future works, we will include more features as the inputs of the prediction algorithm and deeper layers as the hidden layers of the proposed neural network. We will also investigate applying the proposed method for higher-level tasks, such as predicting highly cited papers and authors, and estimating the h-index of the authors in the future.

\bibliographystyle{plainnat}
\bibliography{nncp}

\end{document}